\documentclass[9pt,shortpaper,twoside,web]{ieeecolor2}
\usepackage{generic}
\usepackage{cite}
\usepackage{amsmath,amssymb,amsfonts}
\usepackage{algorithmic}
\usepackage{graphicx}
\usepackage{textcomp}
\usepackage{caption}
\usepackage{subcaption}
\usepackage{siunitx}
\usepackage{float}
\usepackage{physics}
\usepackage{multirow}
\usepackage{hyperref}
\usepackage{array}
\usepackage[normalem]{ulem}
\usepackage{afterpage}

\newcommand*\diff{\mathop{}\!\mathrm{d}}     
\renewcommand{\baselinestretch}{0.948}
\def\BibTeX{{\rm B\kern-.05em{\sc i\kern-.025em b}\kern-.08em
    T\kern-.1667em\lower.7ex\hbox{E}\kern-.125emX}}
\begin{document}
\title{TuneS: Patient-specific model-based  optimization of contact configuration in deep brain stimulation}
\author{Anna F. Frigge \IEEEmembership{Graduate Student Member, IEEE}, Lina Uggla, Elena Jiltsova, Markus Fahlström, Dag Nyholm, and Alexander Medvedev \IEEEmembership{Member, IEEE}
\thanks{This paragraph of the first footnote will contain the date on 
which you submitted your paper for review. This work is funded by the Swedish Research Council via Grant 2020-02901 to the project ``Patient-specific dynamical modeling and optimization of deep brain stimulation" within The EU Joint Programme – Neurodegenerative Disease Research.}
\thanks{A. F. Frigge is with the Department of Information Technology, Uppsala University, Sweden (e-mail: anna.frigge@it.uu.se). }
\thanks{L. Uggla, was with the Department of Information Technology, Uppsala University, Sweden. She is 
now with Stardots AB, Uppsala, Sweden (e-mail: lina.uggla@stardots.se).}
\thanks{E. Jiltsova is with 
the Department of Medical Sciences, Neurosurgery, Uppsala University,  Sweden (e-mail: elena.jiltsova@neuro.uu.se).}
\thanks{M. Fahlström is with 
the Department of Surgical Sciences, Uppsala University, Sweden (e-mail: markus.fahlstrom@uu.se).}
\thanks{D. Nyholm is with 
the Department of Medical Sciences, Neurology, Uppsala University, Sweden (e-mail: dag.nyholm@neuro.uu.se).}
\thanks{A. Medvedev is with 
the Department of Information Technology, Uppsala University, Sweden (e-mail: alexander.medvedev@it.uu.se).}}

\maketitle

\begin{abstract}
\textit{Objective:} The objective of this study is to develop and evaluate a systematic approach to optimize Deep Brain Stimulation (DBS) parameters, addressing the challenge of identifying patient-specific settings and optimal stimulation targets for various neurological and mental disorders.
\textit{Methods:} TuneS, a novel pipeline to predict clinically optimal DBS contact configurations based on predefined targets and constraints, is introduced. The method relies upon patient-specific models of stimulation spread and extends optimization beyond traditional neural structures to include automated, model-based targeting of streamlines.
\textit{Results:} Initial findings show that both the STN motor subdivision and STN motor streamlines are consistently engaged under clinical settings, while regions of avoidance receive minimal stimulation. Given these findings, the value of model-based contact predictions for assessing stimulation targets while observing anatomical constraints is demonstrated at the example of 10 of Parkinson's disease patients. The predicted settings were generally found to achieve higher target coverages while providing a better trade-off between maximizing target coverage and minimizing stimulation of regions associated with side effects.
\textit{Conclusion:} TuneS shows promise as a research tool, enabling systematic assessment of DBS target effectiveness and facilitating constraint-aware optimization of stimulation parameters.
\textit{Significance:} The presented pipeline offers a pathway to improve patient-specific DBS therapies and contributes to the broader understanding of effective DBS targeting strategies.
\end{abstract}


\section{Introduction}
\label{sec:introduction}

Deep Brain Stimulation (DBS) has emerged as a powerful and cost-efficient~\cite{Pietzsch2016} therapeutic tool for addressing a range of neurological and psychiatric disorders. In DBS, stimulation electrodes are implanted into specific areas of the brain to chronically deliver controlled electrical impulses. The precise DBS target area varies depending on the disorder and is selected based on the patient's specific symptoms. A typical DBS target in movement disorders is the subthalamic nucleus (STN) due to its significant role in alleviating symptoms across various disorders with distinct symptoms~\cite{Honey2017,Chandra2022,Ostrem2017,Chabardes2020}.
Despite the effectiveness of DBS in the alleviation of symptoms, optimizing stimulation parameters in clinical practice relies on a trial-and-error method called monopolar review, which involves testing each contact, one at a time, with monopolar stimulation. This process is laborious and time-consuming, often resulting in suboptimal clinical outcomes due to the time constraints faced by clinical staff~\cite{Wagle2017} and wide inter-patient variability of optimal stimulation settings evidenced in numerous studies.
Moreover, the evolution of increasingly sophisticated lead designs has provided clinicians with enhanced control over the distribution of the electric field, while simultaneously increasing the complexity associated with parameter optimization. 

In recent years, there has been an on-going effort to develop automated programming algorithms that can aid clinicians in the programming procedure.
These include image guided algorithms~\cite{Cubo2019,Shub2022}, data-driven approaches utilizing monopolar review data~\cite{Roediger2023}, and algorithms that incorporate quantified symptoms, e.g. tremor measurements via smartwatch applications~\cite{Sarikhani2022}, or other biomarkers like beta activity in local field potentials (LFP)~\cite{Wang2023} or cortical evoked potentials~\cite{Connolly2021}.

Recent studies have shown that image-guided tools, which display the DBS lead relative to key target structures, can reduce initial programming times by providing a useful starting point for clinicians~\cite{Aldred2023}. Additionally, some lead manufacturers offer commercial software for image-guided parameter predictions as part of their DBS systems, which have demonstrated improved clinical outcomes in Parkinson's Disease (PD) patients~\cite{Torres2024}. Shub \textit{et al.}~\cite{Shub2022}   found that predictions made by GuideXT\textsuperscript{\textregistered}, although did not always align with decisions taken by clinical programmers, yet  resulted in significant symptom relief. However, commercial software solutions often provide limited flexibility for individual target selection, constraint customization, and parameter adjustments, which restricts their utility as research tools. Additionally, these platforms typically lack functionality for exporting results to other software for further analysis or processing, thus impeding their integration into broader research workflows.

While data-driven algorithms can be valuable for making predictions, they often fail to provide insights into the underlying mechanisms of DBS. Moreover, these methods frequently struggle with extrapolation, which is particularly challenging in DBS due to the highly individualized nature of the therapy. Additionally, differences in procedures between medical centers, limited data for less common DBS-treated disorders, and the complexity of assessing a patient's overall condition further complicate the use of data-driven approaches. 

Algorithms based on kinematic biomarkers or LFP beta activity are generally applicable only to patients who exhibit these specific biomarkers, limiting their utility.
Additionally, many of these tools primarily focus on motor improvements, potentially overlooking the importance of avoiding stimulation in areas that could cause non-motor side effects.

\begin{figure}[ht]
    \centering
\includegraphics[width=0.8\linewidth]{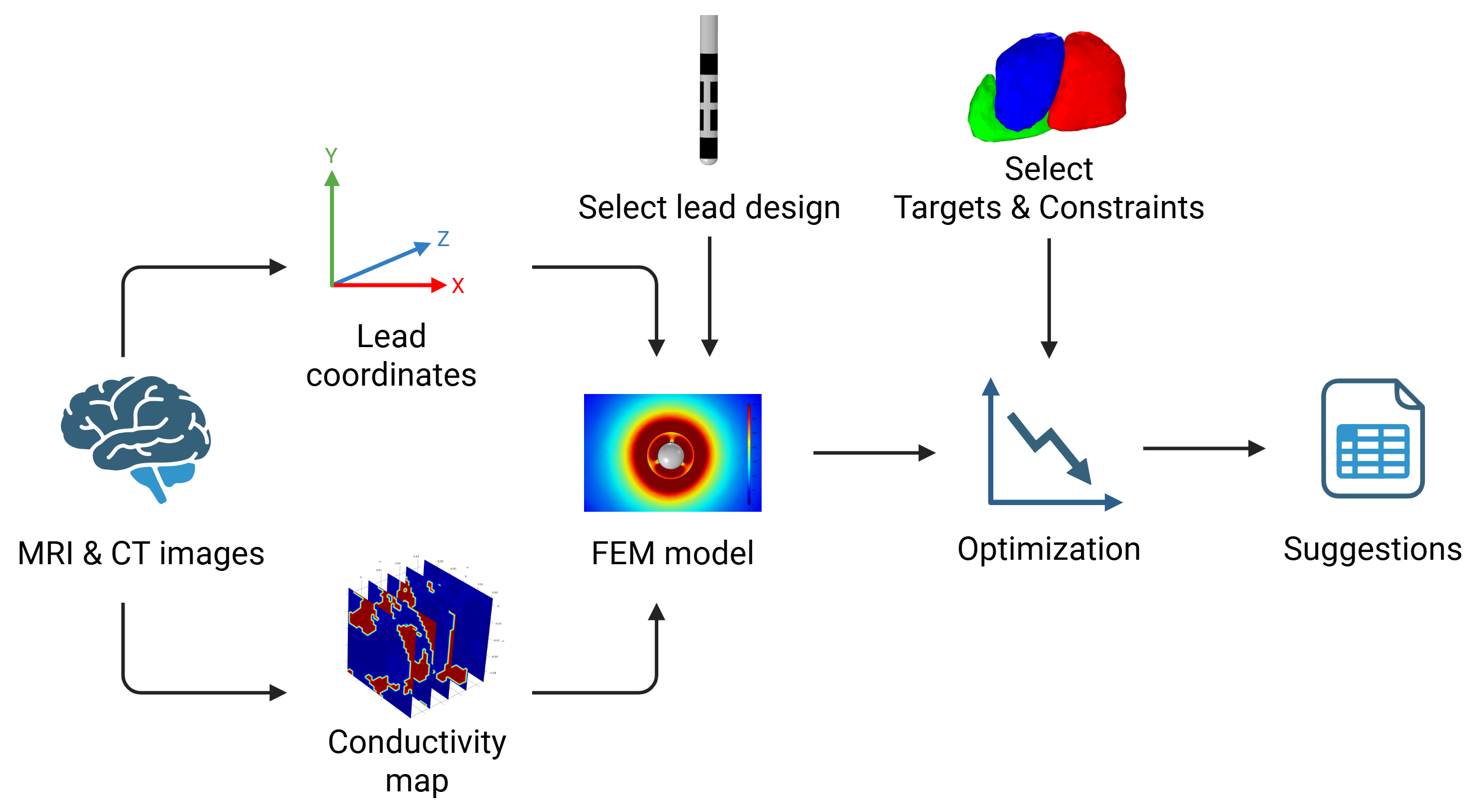}
    \caption{Schematic workflow of TuneS based on preoperative MRI and postoperative CT images~\cite{BioRenderTuneSChart}. 
}\label{fig:TuneSflow}
 
\end{figure}
 
To address these challenges, an automated image-guided programming pipeline called TuneStim (TuneS)~\footnote{TuneS is freely available on \href{https://github.com/annafrigge/TuneStim}{github.com/annafrigge/TuneStim}.} has been developed at Uppsala University in collaboration with Uppsala University Hospital aimed at predicting initial DBS configuration settings that can be further refined by clinical practitioners. Leveraging readily available medical imaging data from routine clinical procedures, such as preoperative magnetic resonance imaging (MRI) and postoperative computed tomography (CT) scans, TuneS facilitates offline computations. The user can select the relevant target structures, including specific anatomical brain regions and reconstructed fibers derived from fiber tractography. To distinguish between actual anatomical fibers and  reconstructions obtained using fiber tractography from Diffusion Tensor Imaging (DTI), the term \textit{streamlines} \cite{Hollunder2024} will be used throughout this paper to specifically refer to the reconstructed tracts. The computed DBS settings can then be seamlessly integrated into patient consultations, streamlining and expediting the programming process. A schematic illustration of the workflow in TuneS is given in Fig.~\ref{fig:TuneSflow}.

The key contributions of the present paper are constituted by a novel open-source DBS software tool as well as a novel stimulation targeting strategy:
\begin{enumerate}
    \item Research tool: TuneS is a freely available software and holds potential as a research tool for investigating relationships between stimulating different DBS targets and constrained regions on the one hand, and therapeutic outcomes and stimulation-induced side-effects on the other hand. 
    \item Streamline targeting: TuneS offers the option to target streamlines, rather than solely relying  on traditional anatomical targets. This is in line with previous reports of automated DBS programming routines including streamline targeting~\cite{Anderson2018,Pena2018,Hines2024,Vorwerk2019}. TuneS can be employed with either atlas-based targets or targets reconstructed from individual patient imaging, thereby making optimization-based predictions more accessible to the research community. 
    Here,  the difference in a point-wise versus a trajectory-wise approach when quantifying streamline activation is demonstrated. 
\end{enumerate}
Besides PD, DBS is effective in treating other neurological disorders, e.g.  essential tremor, dystonia, drug-resistant epilepsy, as well as various psychiatric disorders e.g. obsessive-compulsive disorder, depression, Tourette syndrome and other conditions.

While TuneS is capable of evaluating stimulation with respect to different targets and predicting optimal contact configurations in all the disorders treated with DBS, the findings presented in this paper solely pertain to PD patients with STN DBS. 
To illustrate the  steps of DBS settings optimization with TuneS and validate the developed software, a cohort  of ten PD patients treated at Uppsala University Hospital is considered.

\section{Methodology}
\subsection{Patient cohort}
\begin{figure}
    \centering
    \includegraphics[width=2.7cm]{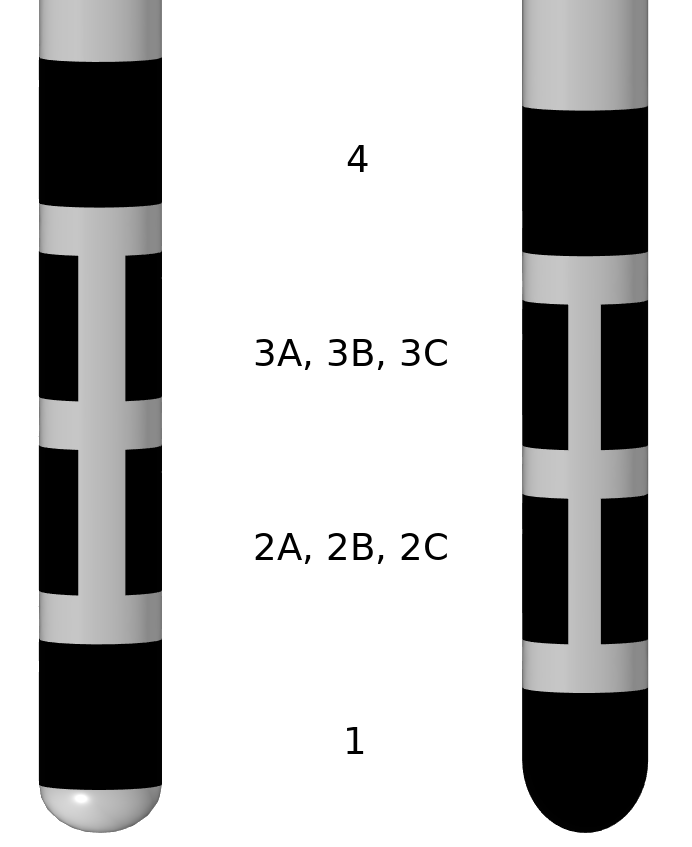}
    \caption{Schematic illustration of the Boston Vercise Cartesia\textsuperscript{\texttrademark} Directional Lead (left) and the Abbott's St. Jude Medical Infinity\textsuperscript{\texttrademark} Directional Lead (right) with the contact labels used in this study. Note that the segmented contacts (A, B, C) are labeled in a clockwise direction.}
    \label{fig:leads}
\end{figure}

The workflow of the TuneS is validated at a cohort of ten Parkinson's disease patients who underwent DBS surgery at Uppsala University Hospital\footnote{The study was approved by the Swedish Ethics Review Authority, Registration number 
2019–05718, and all participants gave informed written consent prior to the beginning of the study.}. The STN was selected as the surgical target for all patients, with nine receiving bilateral DBS and one undergoing unilateral DBS, resulting in a total of 19 leads. Half of the patients were implanted with the Boston Vercise Cartesia\textsuperscript{\texttrademark} Directional Lead, while the other half were implanted with the Abbot's St. Jude Medical Infinity\textsuperscript{\texttrademark} Directional Lead as depicted in Fig.~\ref{fig:leads}.

For each patient, preoperative T1- and T2-weighted MRI, as well as post-operative CT images acquired during routine clinical procedures, were obtained. 
Although the timing of the CT scans post-surgery varied between patients, all images were visually checked for major air pockets that could potentially affect the reconstruction of the lead location relative to the stimulation targets.

\subsection{Image processing}
Pre-operative MRI (T1 and T2) as well as a post-operative CT images were co-registered and normalized using Lead DBS v2.6~\cite{Horn2019}. The lead coordinates were then reconstructed using the PaCER algorithm~\cite{Husch2018}. Moreover, the DiODE algorithm~\cite{Dembek2021_DiODe} was applied to obtain the lead orientation based on the marker artifacts in the post-operative CT. While the latter algorithm has been validated for the Boston Vercise Cartesia\textsuperscript{\texttrademark} Directional Lead, it has not been validated for the Abbott's St. Jude Medical Infinity\textsuperscript{\texttrademark} Directional Lead. 

\subsection{TuneS}
\begin{figure}[ht]
    \centering
    \includegraphics[width=0.7\linewidth]{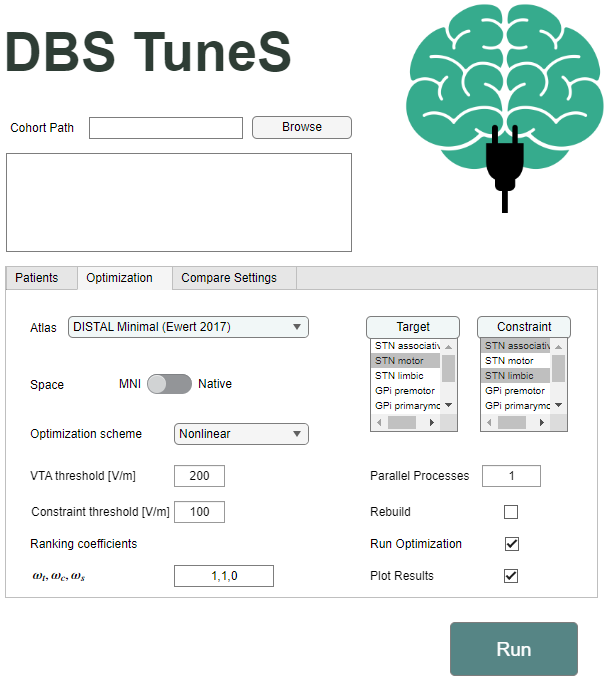}
    \caption{Graphical user interface of TuneS, enabling patient cohort processing, individual target and constraint selection, and model parameter modification. }
    \label{fig:TuneSGui}
\end{figure}

TuneS features a graphical interface as illustrated in Fig.~\ref{fig:TuneSGui}. Implemented in MATLAB, the pipeline relies on dependencies including SPM12, Lead-DBS v2.6 ~\cite{Horn2019} or v3.0~\cite{Neudorfer2023}, and COMSOL Multiphysics\textsuperscript{\textregistered} with Livelink\textsuperscript{\texttrademark} for MATLAB. 

\paragraph{FEM model}
The FEM model is built in COMSOL Multiphysics\textsuperscript{\textregistered} using LiveLink\textsuperscript{\texttrademark} for MATLAB.
Currently, TuneS features four lead designs, including the Abbot's (St. Jude) Medical Infinity\textsuperscript{\texttrademark} Directional lead, the Boston Scientific Vercise Standard lead, the Boston Scientific Vercise Cartesia\textsuperscript{\texttrademark} Directional lead, and the Medtronic 3887 lead.

Surrounding the chosen lead type, heterogeneous tissue is modeled in a $\SI{50}{mm} \times \SI{50}{mm} \times \SI{50}{mm}$ box. To account for the heterogeneous tissue properties in vicinity of the lead, a conductivity map is generated based on a segmentation of the $\SI{7}{T}$ ICBM 152 2009a Nonlinear Asymmetric T1 template MRI~\cite{Fonov2011}. Further away from the lead, the tissue is modeled as homogeneous matter. The default conductivity values for gray matter (GM), white matter (WM), and cerebrospinal fluid (CSF), the encapsulation layer, and the homogeneous medium are $\sigma_{\mathrm{GM}} = \SI{0.09}{S/m}$~\cite{Andreuccetti2017}, $\sigma_{\mathrm{WM}}= \SI{0.06}{S/m}$~\cite{Andreuccetti2017}, $\sigma_{\mathrm{CSF}}= \SI{2.0}{S/m}$~\cite{Andreuccetti2017},  $\sigma_{\mathrm{enc}}= \SI{0.18}{S/m}$~\cite{Chaturvedi2010}, and $\sigma_{\mathrm{hom}}= \SI{0.1}{S/m}$, respectively. In TuneS, these values can be changed by the user. The electrode contacts are treated as perfect conductors, whereas the rest of the lead is assumed to be perfectly insulated.

The FEM model solves the quasi-static approximation of Maxwell's equation
\begin{equation}
    \nabla \cdot (\sigma \nabla u) = 0,
    \label{eq:pde_static}
\end{equation}
where $\nabla \cdot$ is the divergence operator,  $\sigma$ denotes the conductivity, and $\nabla u$ stands for the gradient of the electric potential. All non-active contact surfaces were given a floating boundary condition.

The boundary condition on the active contacts is given by the surface integral
\begin{equation}
    \int \limits_{\partial\Omega} \mathbf{J}\cdot \mathbf{n} \diff S = I_0,
    \label{eq:boundary_condition}
\end{equation}
where $\mathbf{J}$ represents the current density, $\mathbf{n}$ is the normal vector to the contact surface, and $I_0$ is the constant current amplitude.
The FEM model comprises approximately $1,000,000$ elements, with slight variations depending on the lead type. The element size near the DBS lead surface is well below the lead radius and increases with distance from the lead as shown in Fig.~S1 in the supplementary material.

\paragraph{Optimization schemes}
The effect of DBS on the neural tissue surrounding the lead is typically approximated by the volume of tissue activated (VTA). 
A point is assumed activated and therefore included in the VTA, if the electric field norm $E$ at this point exceeds a specified threshold $E_{\mathrm{th,t}}$. Exploiting the linearity of~\eqref{eq:pde_static}, the field obtained from the FEM model under a unit stimulus can be linearly scaled to obtain the results for different amplitudes.
It is noteworthy that TuneS employs point clouds in calculating VTA rather than volumes. Targets and constraints are represented as disjoint sets, and when multiple targets or constraints are involved, the union of these disjoint sets defines the target volume $\Omega_{\mathrm{t}}$ and the constraint volume $\Omega_{c}$, respectively. In the following, the 3D point coordinates $(x,y,z)$ of each point in $\Omega_\mathrm{t}$ and $\Omega_\mathrm{c}$ are mapped to the indices $i$ and $j$, respectively, using a coordinate-to-index conversion. In this mapping, the innermost loop iterates through $x$, followed by $y$, and finally $z$. To decrease computational load, the points in the target and constraint volumes can be filtered using a voxel filter, which downsamples by keeping only one representative point within each voxel of a specified size. Here, the voxel filter size was set to \SI{0.9}{mm}.

Based on the above concept of tissue activation, the current version of the TuneS features a set of optimization schemes, two of which are presented in the following. 

The first one is a linear optimization scheme that aims to activate as many of the target points as possible while keeping the  scaled electric field norm $\lambda \cdot E_\mathrm{j}$ at $\theta \ \%$ of points $j$ within the constraint volume $\Omega_\mathrm{c}$ under a specific constraint threshold value $E_{\mathrm{th,c}}$. It can be mathematically formulated as 
\begin{equation}
\begin{aligned}
  \lambda_{\mathrm{opt} }=\arg\max_\lambda \quad & \omega_t \cdot\sum_{i \in \Omega_t } \lambda \cdot E_i - \omega_c \cdot \sum_{j \in \Omega_c} \lambda \cdot E_j,\\
    \text{s.t.} \quad &  \lambda \cdot E_\mathrm{j} \leq E_{\mathrm{th}}\quad \text{for } \theta \ \% \text{ of points $j$ in } \Omega_{c} , \\
    & \lambda \geq 0
\end{aligned}
\label{eq:opti1}
\end{equation}
where $\lambda$ is the stimulation amplitude for a
given combination of active contacts, assuming a uniform distribution of
current through all contacts and $E_i$ denotes the electric field norm at the points $i$ within the target volume $\Omega_\mathrm{t}$. 
\begin{figure}[t!]
    \centering
    \begin{subfigure}[b]{0.49\columnwidth}
    \includegraphics[height=4.6cm]{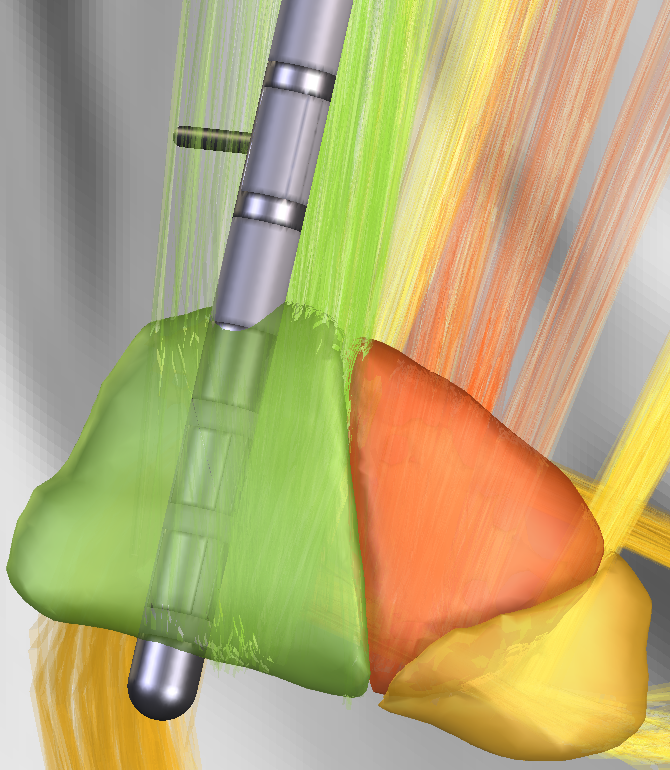}
    \caption{}
    \end{subfigure}
    \begin{subfigure}[b]{0.49\columnwidth}
        \includegraphics[height=4.6cm]{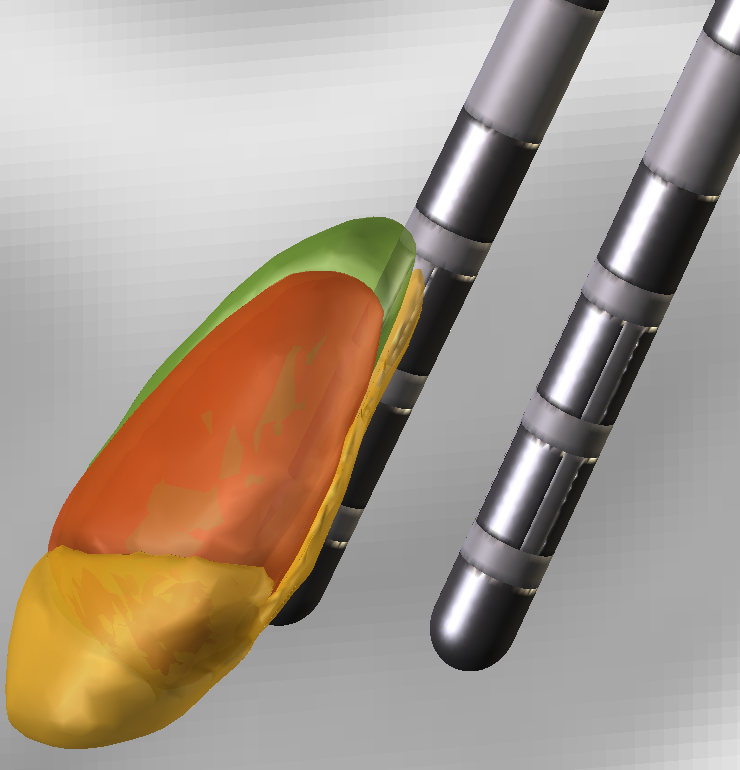}
        \caption{}
        \end{subfigure}
    \caption{(a) DBS lead location relative to STN subdivisions (motor - green, associative - orange, limbic - yellow). (b) Reconstruction of the left lead in Patient 10 based on CT scans acquired three days post-surgery (lead at the border of the STN) and several months later (lead further from the STN). The apparent displacement of the lead relative to the STN subdivisions can be attributed to brain shift effects.}
    \label{fig:pat10}
\end{figure}
The second optimization scheme uses a nonlinear cost function that penalizes under-stimulation of target points quadratically, while overstimulation is penalized linearly, as proposed in~\cite{Cubo2015}. It is given by

\begin{equation}
   f_{\mathrm{cost,}i} = 
    \begin{cases}
      \abs{\lambda \cdot E_i - E_{\mathrm{th,t}}}^2  & \text{if } \lambda\cdot E_i \leq  E_{\mathrm{th,t}}\\
      \abs{\lambda \cdot E_i - E_{\mathrm{th,t}}}  &  \text{if } \lambda \cdot E_i > E_{\mathrm{th,t}}
    \end{cases}       
    , 
\end{equation}
where $E_{\mathrm{th,t}}$ denotes the desired target field strength and $\abs{\cdot}$ represents the absolute value. Minimization of the cost function including a linear constraint avoidance penalty gives
\begin{align}\label{eq:opti2}
   \lambda_{\mathrm{opt}}=\arg \min_\lambda \quad & \omega_t \cdot \sum_{i \in \Omega_{\mathrm{t}}} f_{\mathrm{cost},i} - \omega_c \cdot \sum_{j \in \Omega_c} \lambda \cdot E_j\\
    \text{s.t.} \quad & \lambda \cdot E_j \leq E_{\mathrm{th,c}}\quad \text{for } \theta \ \% \text{ of points $j$ in } \Omega_{c}. \nonumber \\
    & \lambda \geq 0 \nonumber
\end{align}
In the case of the eight-contact leads illustrated in Fig.~\ref{fig:leads}, these optimization problems are solved for 31 different contact combinations, including individual contacts, adjacent segmented contacts, ring-segment pairings, three-segment rings, and vertically misaligned combinations. The configurations represent a clinically relevant subset of all
possible configurations and allow to reduce the necessary computations. The choice of the activation threshold $E_{\mathrm{th,t}}$ is strongly dependent on the stimulation pulse width and the axon diameter of neurons in the tissue surround the DBS lead~\cite{Astrom2015}. In this paper, it is assumed that $E_{\mathrm{th,c}} = \SI{100}{S/m \ (1.7)}$ and  $E_{\mathrm{th,t}} = \SI{200}{S/m \ }$. In the following, a change in the constraint variable $\theta$ is mainly handled by introducing a constraint relaxation variable $\gamma$. Specifically, a relaxation of the constraint variable $\theta$ by $\gamma$ indicates that the constraints in~\eqref{eq:opti1} and~\eqref{eq:opti2} are satisfied for $\theta = 100 - \gamma$ percent of the points in $\Omega_c$. 

Optimization problem~\eqref{eq:opti1} is solved using MATLAB’s {\tt linprog}, whereas the minimization of the cost function in \eqref{eq:opti2} is handled by {\tt fmincon}.

\paragraph{Targets and constraints}
The selection of stimulation target and constraint areas is left to the user as it may vary between different disorders and is subject of on-going research. In this paper,  TuneS was employed to investigate two potential target structures in DBS for PD: (1) the subdivisions of the STN~\cite{Ewert2018} and (2) the STN streamlines~\cite{Middlebrooks2020}, specifically the reconstructed hyperdirect pathways connecting STN subdivisions to corresponding cortical regions. Fig.~\ref{fig:pat10} illustrates the STN subdivisions and streamlines, which were here derived from averaged patient cohort atlases, which may not fully capture patient-specific anatomy. 
The STN motor (respectively the STN motor streamlines) was considered as the stimulation target, while the STN limbic and STN associative regions (respectively their streamlines) were delineated as constraint areas.

\begin{figure}[t]
    \centering
    \includegraphics[width=0.8\columnwidth]{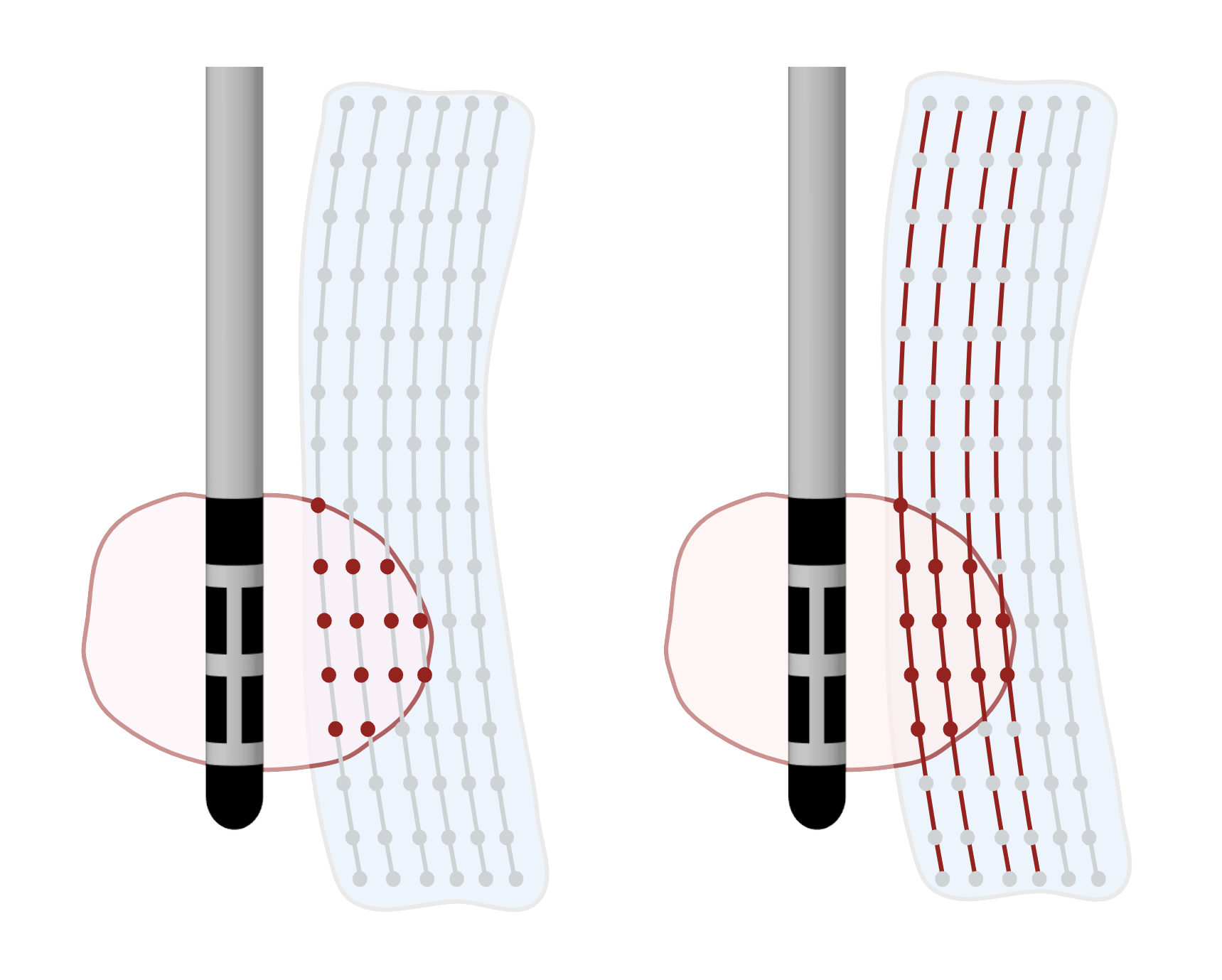}
    \caption{Point-wise (left) and trajectory-wise (right) quantification of tract activation in a static model. Red points indicate locations where the electric field norm exceeds the activation threshold. In the trajectory-wise approach, entire trajectories are marked in red when at least one point along their course surpasses the threshold.}
    \label{fig:tract_activation}
\end{figure}

Generally, STN streamlines consist of a much higher number of points, which span from the STN region all the way to the motor cortex. By removing points outside of the region of interest i.e. in the vicinity of the DBS lead, the number of points is reduced. Nevertheless, when target activation is quantified point-wise, the resulting coverage percentages can still be relatively low, depending on the size of the region of interest. Alternatively, it can be assumed that the entire trajectory of the streamline is activated if any point along its path exceeds a specific field threshold. An illustration of the two approaches is given in Fig.~\ref{fig:tract_activation}.
In order to incorporate the trajectory-wise activation into the optimization scheme, a pre-selection of target and constraint points is performed. Given the linearity of~\eqref{eq:pde_static}, the scaling of the electric field norm can be reduced to the point on each trajectory that exhibits the maximum field norm under a unit stimulus. Accordingly, for each trajectory, only this point was retained for inclusion in the optimization, both for target and for constraint streamlines.

\label{chap:2TargetsAndConstraints}

\paragraph{Configuration suggestions}
The performance of active contact configurations can be ranked in various ways. In this study, the optimized settings for all contact configurations were ranked by a heuristic score $S$, defined as 

\begin{equation}
    S = \omega_\mathrm{t} \cdot p_{\mathrm{act,t}} - \omega_\mathrm{c} \cdot p_{\mathrm{act,c}} - \omega_\mathrm{s} \cdot p_{\mathrm{act,s}},
    \label{eq:score}
\end{equation}

where $p_{\mathrm{act,t}}$ and $p_{\mathrm{act,c}}$ represent the percentage of target and constraint activated, respectively. Additionally, $p_{\mathrm{act,s}}$ denotes the percentage spill, i.e. the percentage of activated tissue outside the target region. The weights, $\omega_t$, $\omega_c$, and $\omega_s$ can be defined by the user from clinical preferences. 
When targeting streamlines, defining spill is a rather complex task due to void space between individual streamlines. Therefore, in this study, it was assumed that $\omega_\mathrm{t} = 1$, \textcolor{blue}{$\omega_\mathrm{c} = 2$}, and $\omega_\mathrm{s} = 0$. 
While exploring different ratios of these ranking coefficients is an important topic, it lies beyond the scope of the present paper. 

\paragraph{Clinical configurations}
When comparing the contact configurations suggested by TuneS to those used clinically, it is important to note that only a limited number of combinations can be tested in clinical practice. This is primarily due to time constraints faced by clinical staff and the variable time it takes for patients to experience the effects of a given setting, with some side effects potentially only emerging after several hours or days. Therefore, the clinically used configurations do not necessarily represent ground-truth in optimization, but rather serve as a reference point. A stimulation setting used clinically is also typically free of adverse side effects, which property corresponds, with respect to the considered optimization problems, to low values of constraint coverage.

\section{Results}
All the computations have been performed in MNI space to facilitate inter-subject comparability. 

\subsection{Clinical Settings}
\begin{figure}[t]
    \centering
    \includegraphics[width=0.9\linewidth]{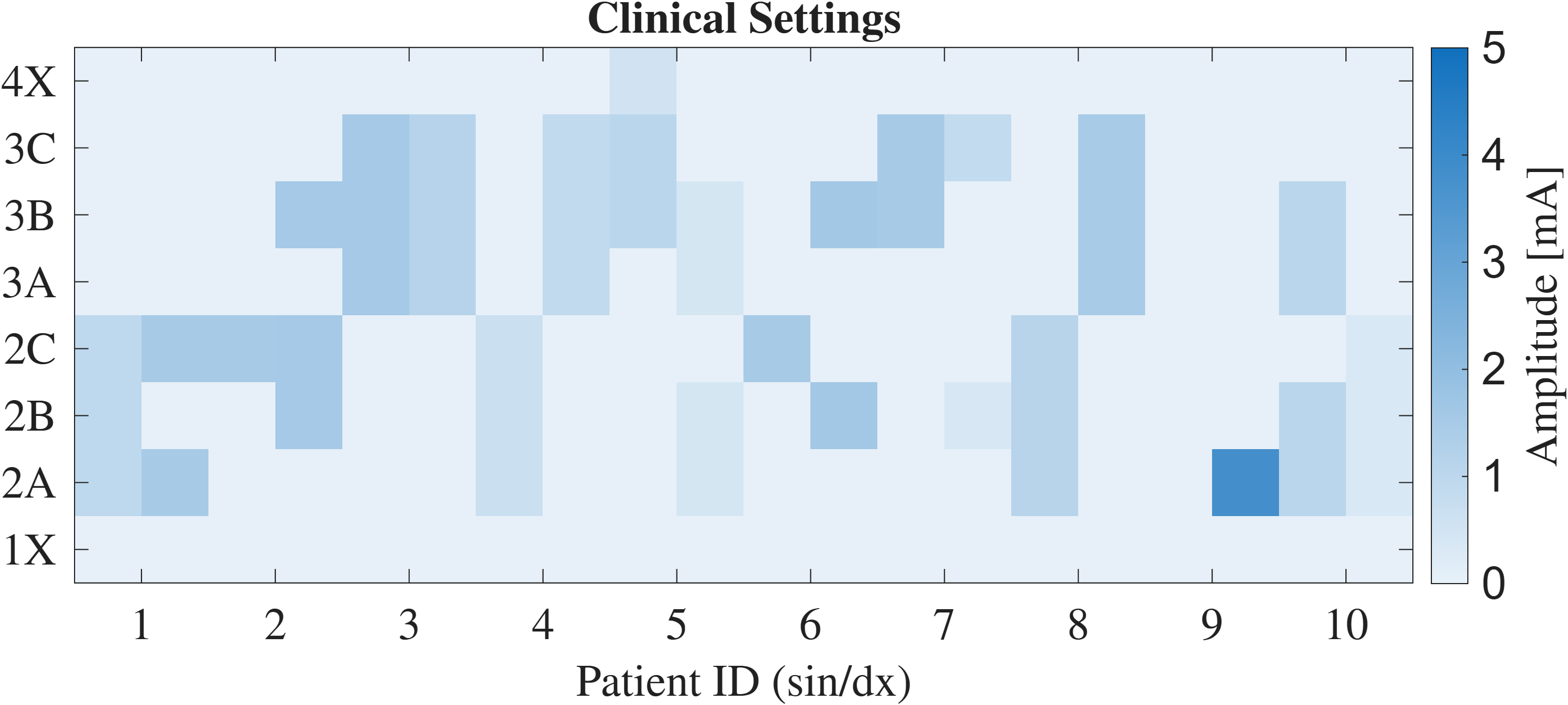}
    \caption{Clinically active contact configurations and current amplitudes for all patients. Current amplitude is color-encoded from \SI{0}{mA} (white) to \SI{5}{mA} (blue). Sin and dx denote left and right hemispheres, respectively. Notably, segmented contacts (A, B, C) are frequently activated together rather than individually. }
    \label{fig:clinical}
\end{figure}

The most recent clinically used settings for all patients, including the active contacts and stimulation amplitude, are given in Fig.~\ref{fig:clinical}. 

\begin{figure}[t]
    \centering
    \begin{subfigure}[b]{0.99\columnwidth}
        \centering
        \includegraphics[width=\textwidth]{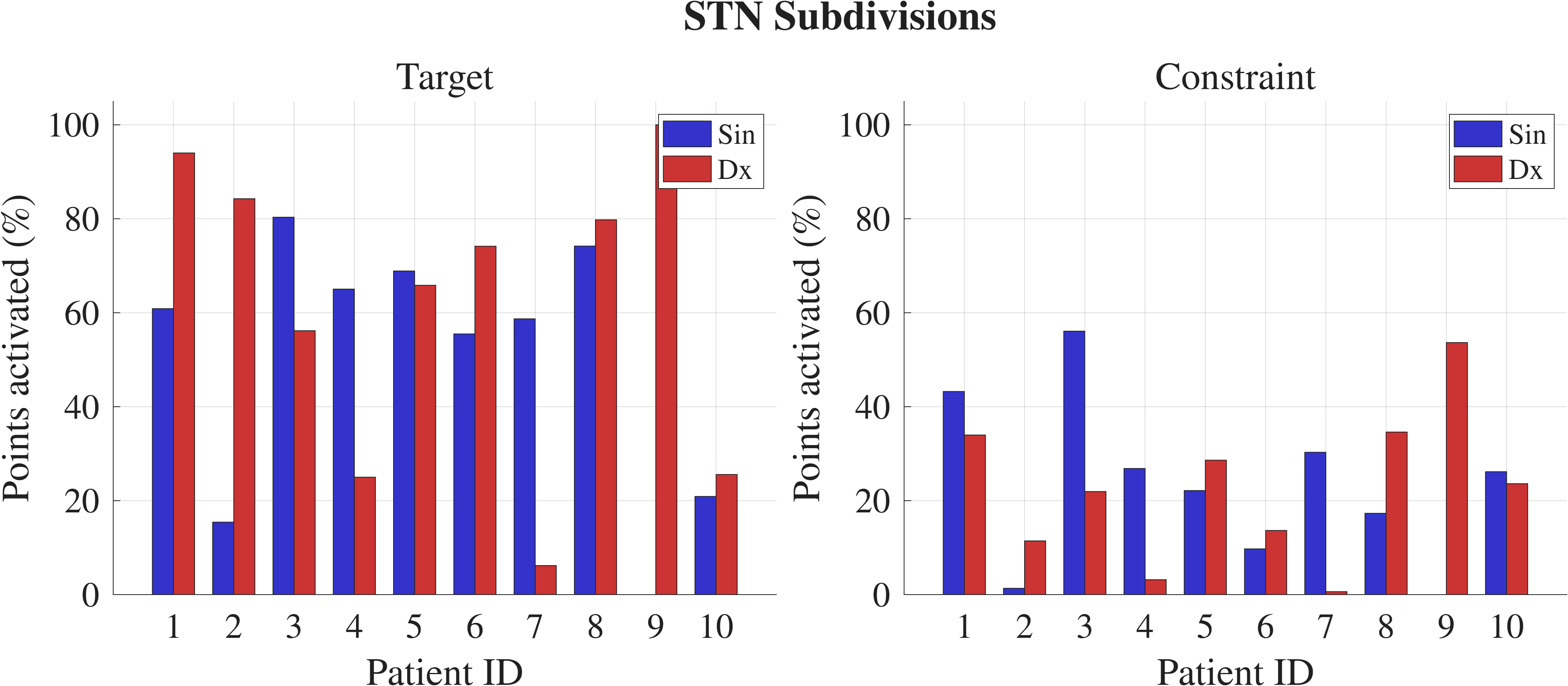}
        \label{fig:coverages1}
    \end{subfigure}
        \begin{subfigure}[b]{0.99\columnwidth}
        \centering
        \includegraphics[width=\textwidth]{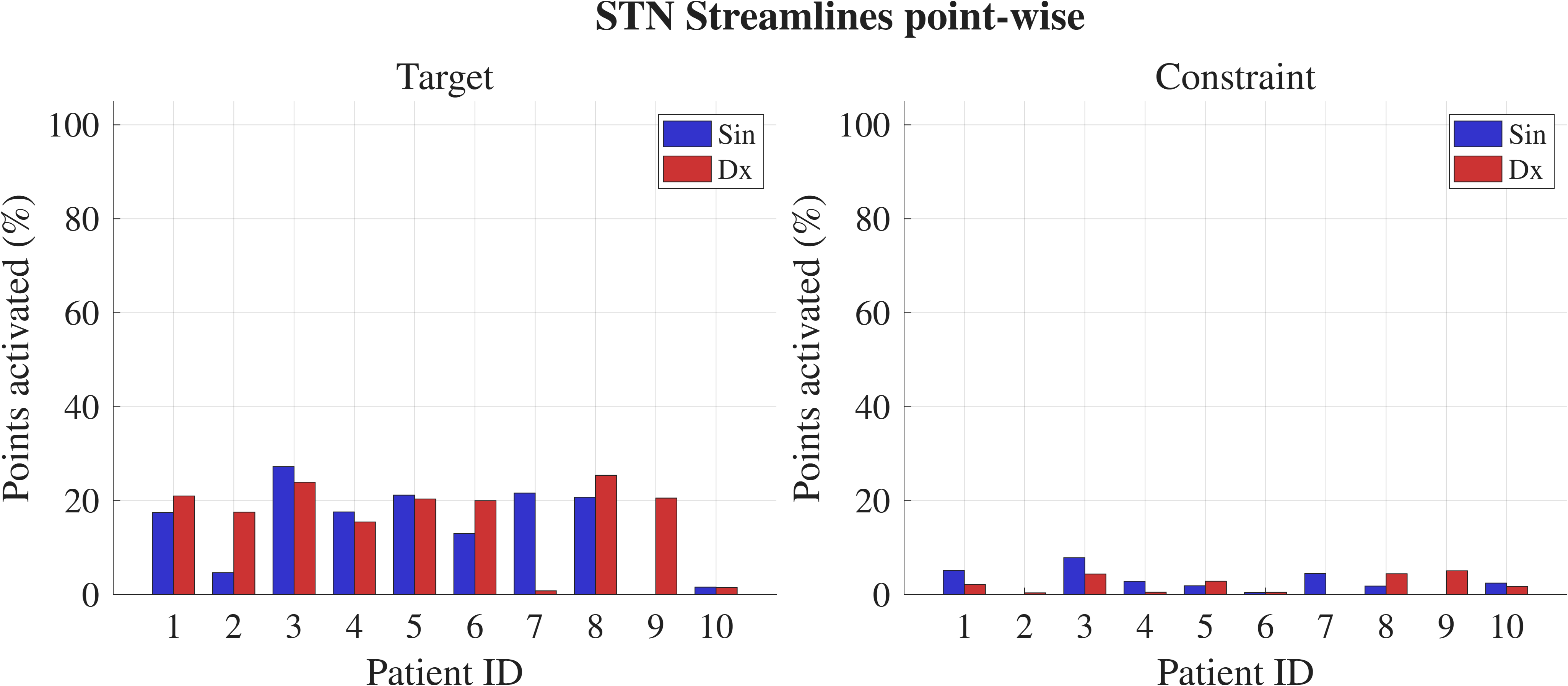}
        \label{fig:coverages2}
    \end{subfigure}
    \begin{subfigure}[b]{0.99\columnwidth}
        \centering
        \includegraphics[width=\textwidth]{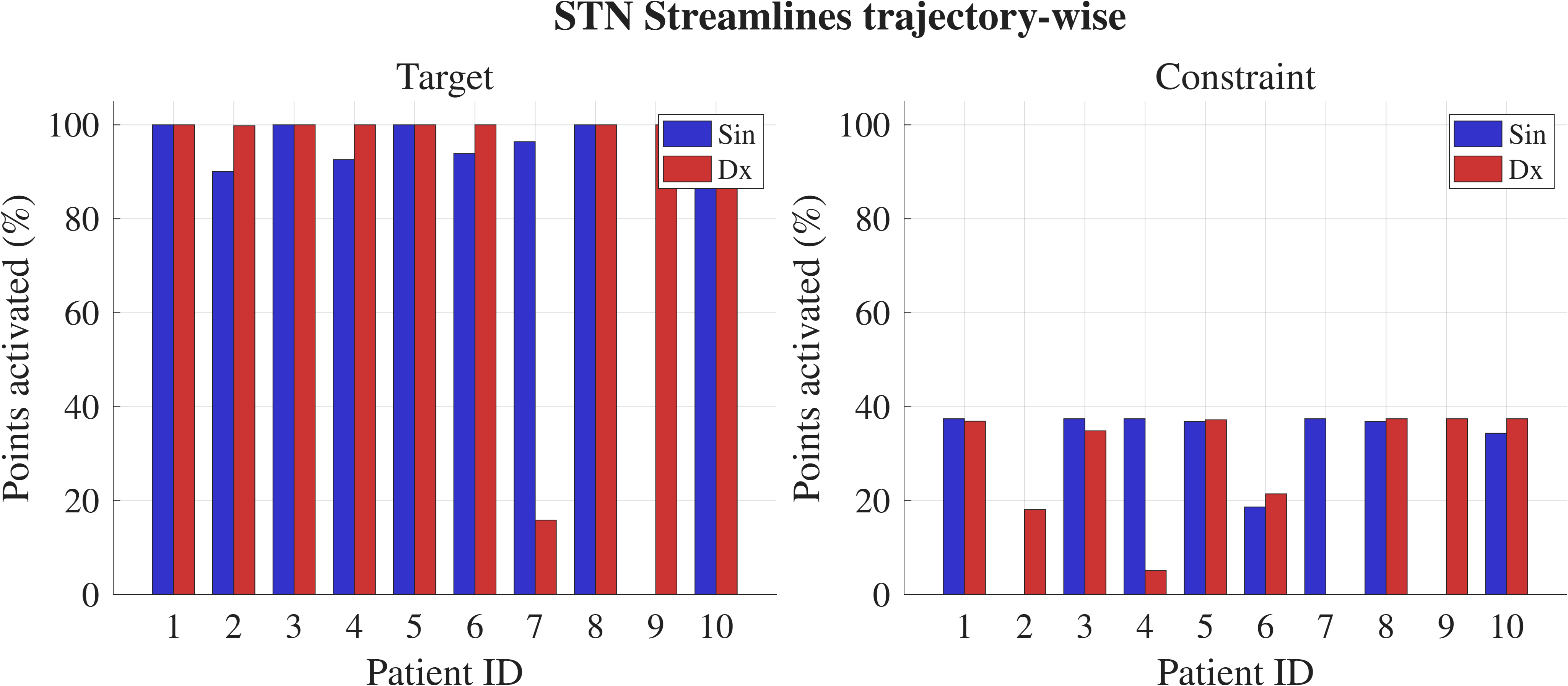} 
        \label{fig:coverages3}
    \end{subfigure}
    
    \caption{Target and constraint coverage under clinical settings for the STN subdivisions, point-wise activation of the STN streamlines, and trajectory-wise streamline activation of the STN streamlines. }     
    \label{fig:coverages}
\end{figure}

The target and constraint coverage for both the STN subdivisions and STN streamlines under these clinical settings were computed from the individualized FEM models and are presented in Fig.~\ref{fig:coverages}. The VTA threshold was adjusted according to the pulse width used clinically, following~\cite{Astrom2015}. For the STN subdivisions, low target coverage was observed from the left lead of Patient~2 and the right lead in Patient~7. In Patient~3, both high target coverage and relatively high constraint coverage were noted, likely due to the higher stimulation amplitude used for this patient ($\SI{4.6}{mA}$). 
In the second row of Fig.~\ref{fig:coverages}, results from a point-wise approach for streamline activation are presented, where only the portion of the streamlines in the region near the lead were considered and further treated as a point cloud. This method leads to relatively small target coverage percentages due to the large size of the region of interest. The last row Fig.~\ref{fig:coverages} illustrates the results, when streamline activation is instead quantified in a trajectory-wise manner. 
All but one of the clinical settings achieved from 90 to $\SI{100}{\%}$ target coverage and up to $\SI{40}{\%}$ of constraint coverage.

\begin{figure}[th]
    \centering
    \begin{minipage}[t]{0.495\columnwidth}
        \centering
        \textbf{Linear optimization scheme}\\[0.5em]
        \begin{subfigure}[t]{\linewidth}
            \includegraphics[width=\linewidth]{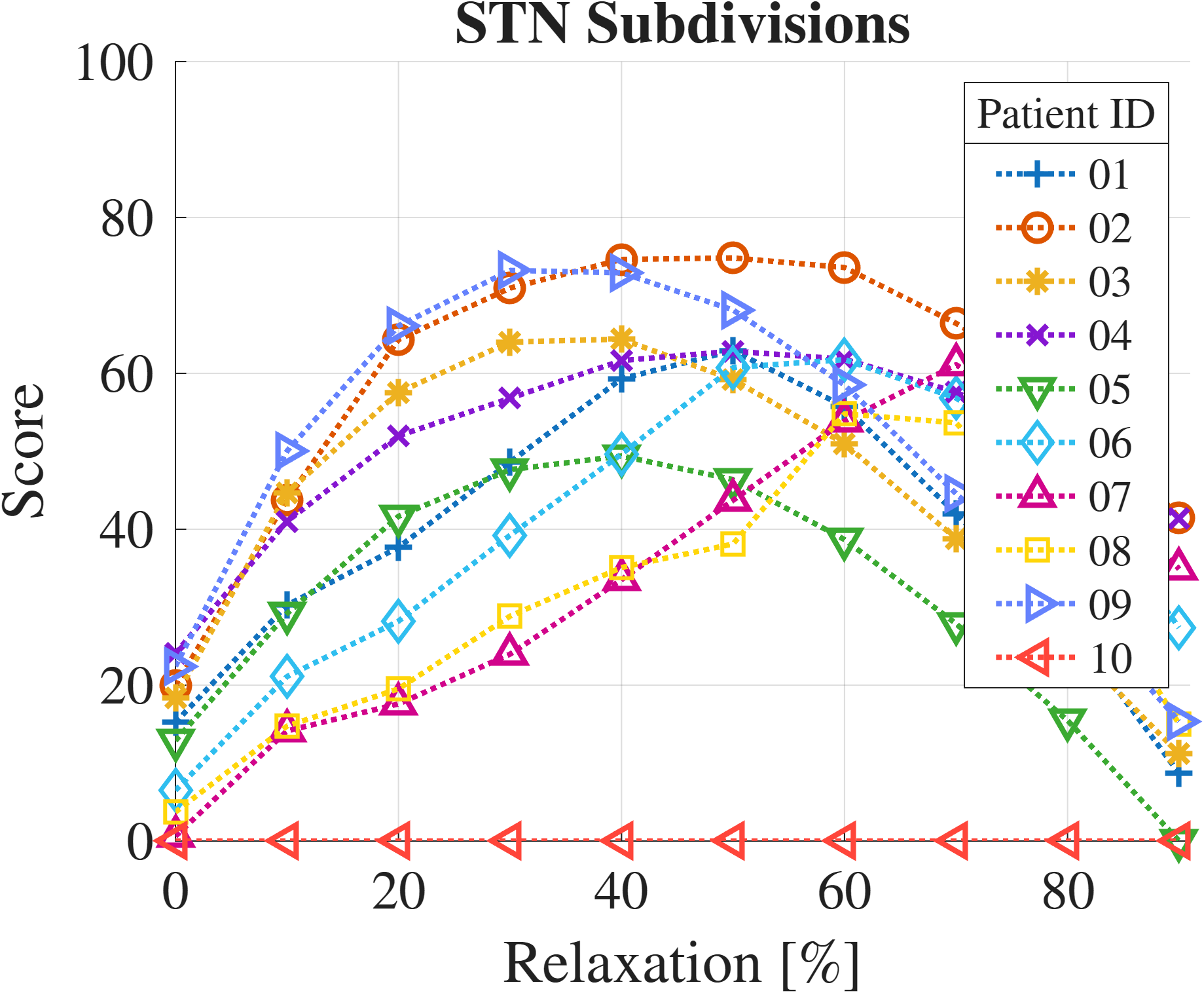}
        \end{subfigure}\vspace{0.5em}

        \begin{subfigure}[t]{\linewidth}
            \includegraphics[width=\linewidth]{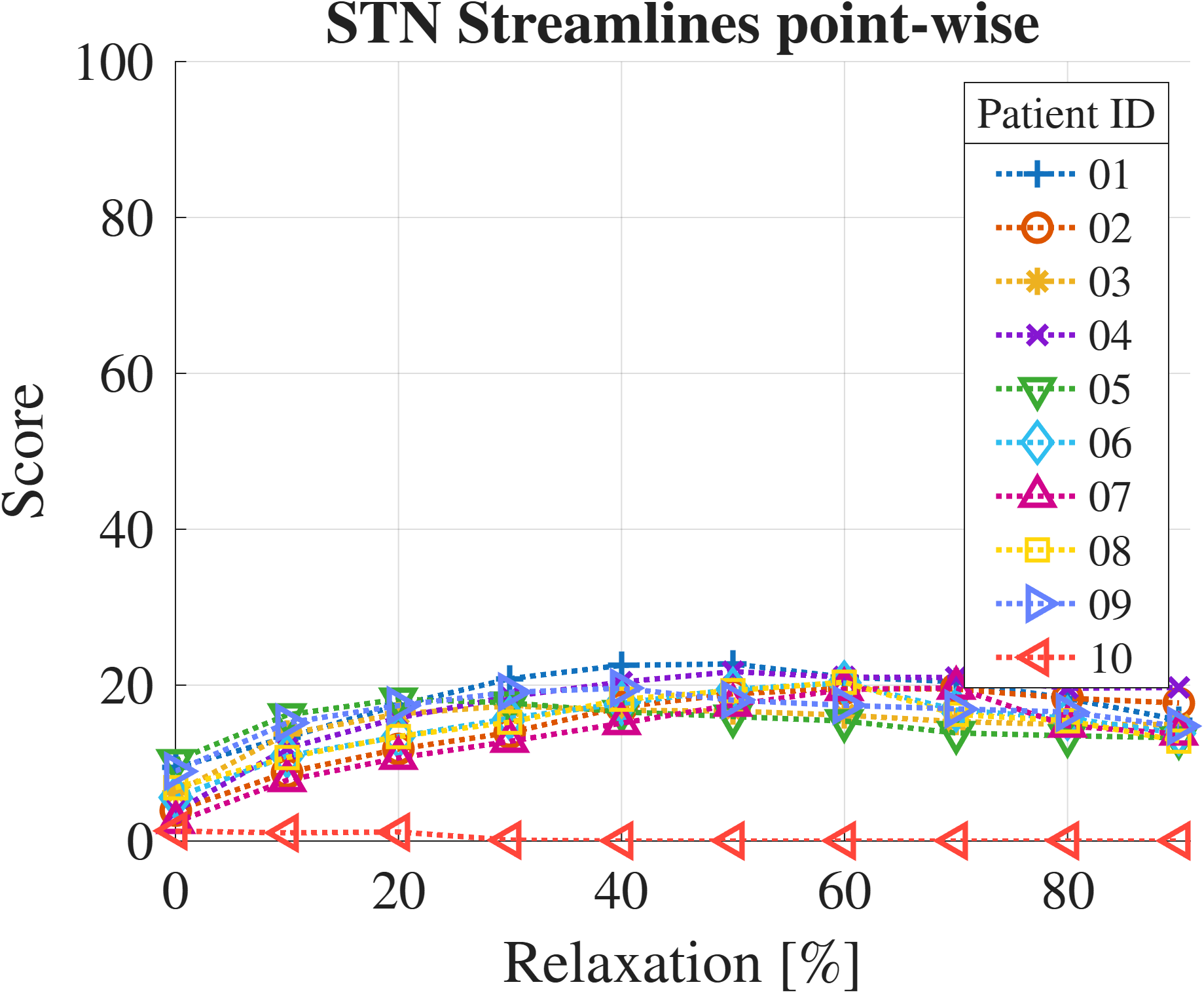}
        \end{subfigure}\vspace{0.5em}

        \begin{subfigure}[t]{\linewidth}
            \includegraphics[width=\linewidth]{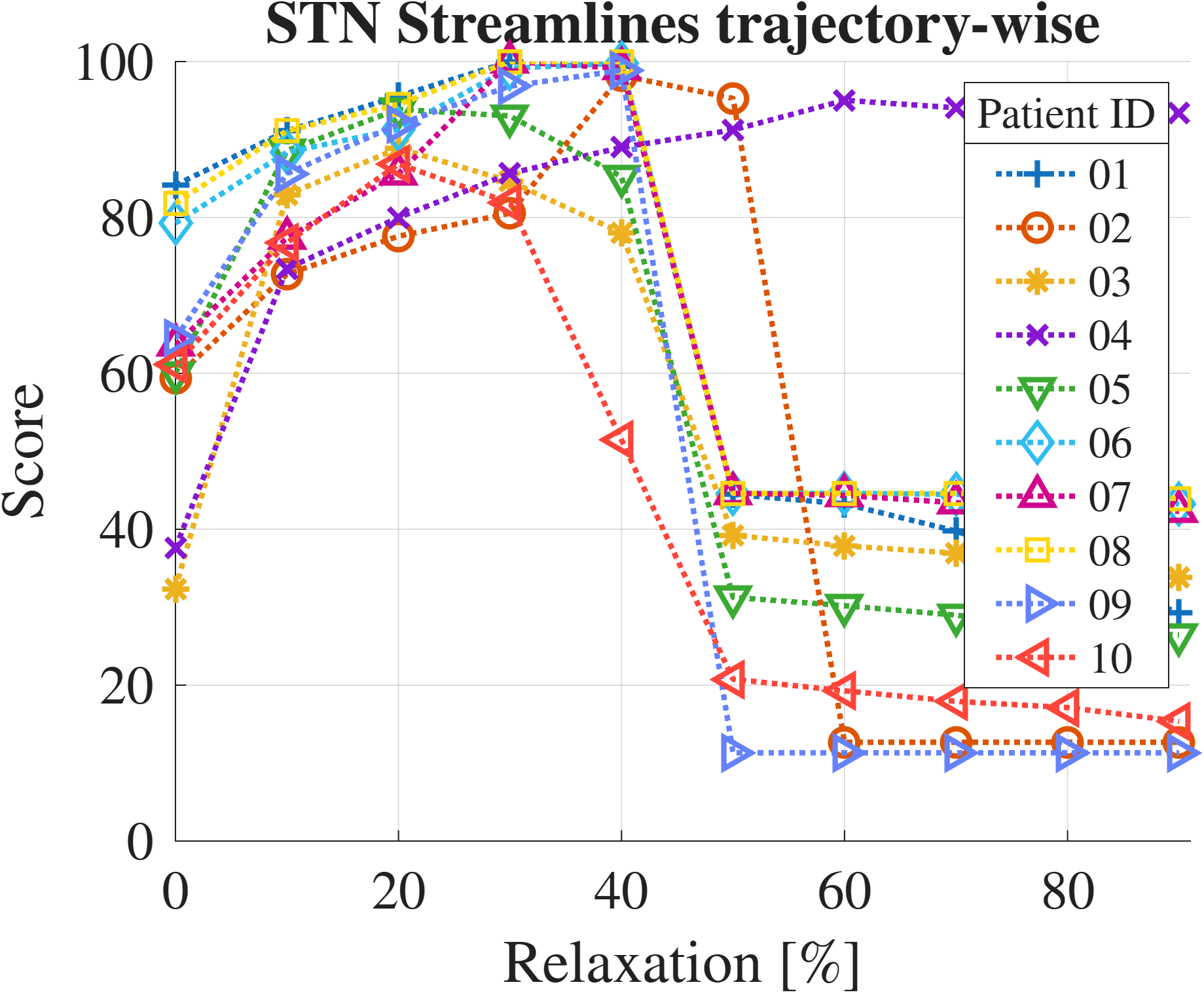}
        \end{subfigure}
    \end{minipage}%
    \hfill
    \begin{minipage}[t]{0.495\columnwidth}
        \centering
        \textbf{Nonlinear optimization scheme}\\[0.5em]
        \begin{subfigure}[t]{\linewidth}
            \includegraphics[width=\linewidth]{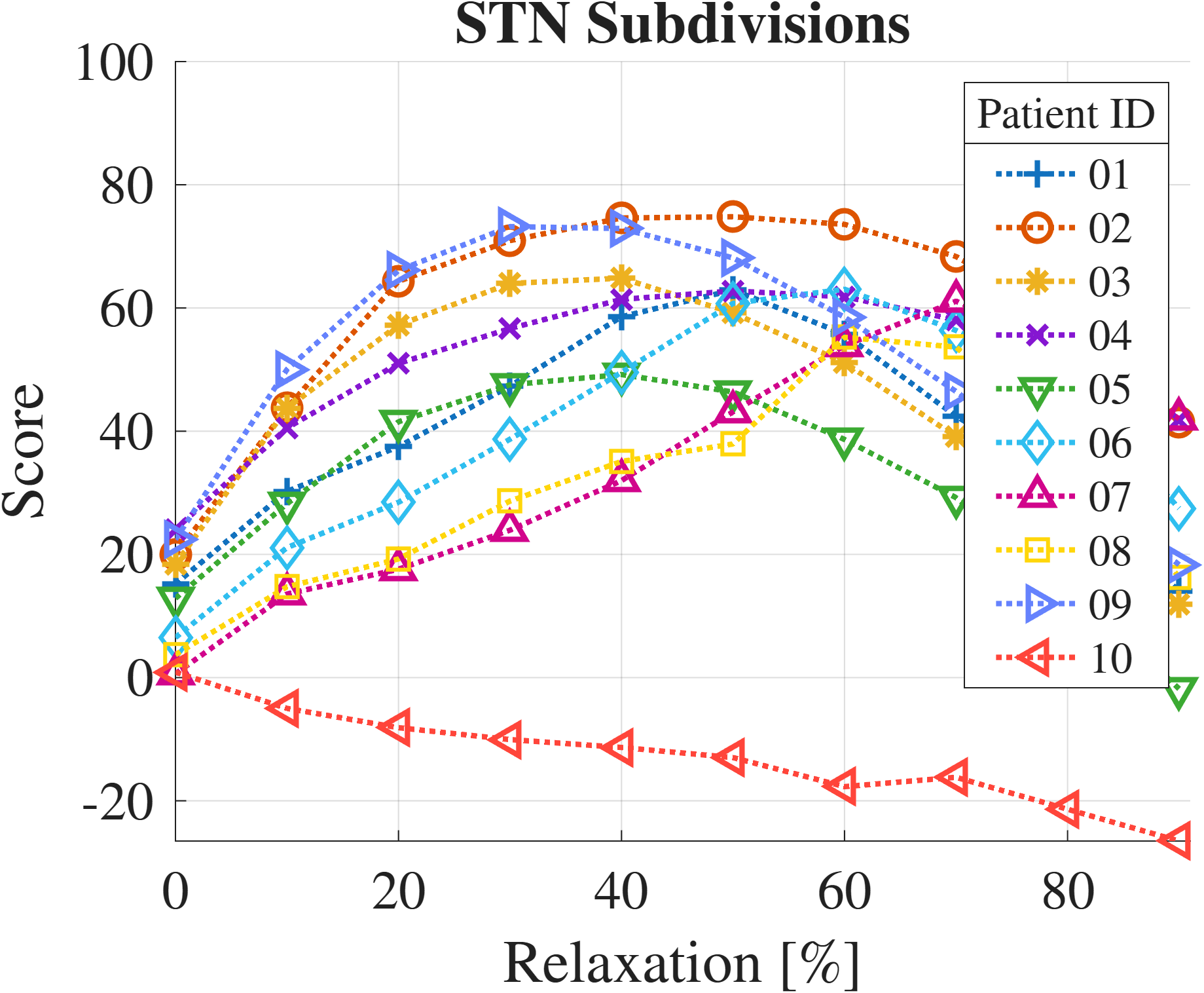}
        \end{subfigure}\vspace{0.5em}

        \begin{subfigure}[t]{\linewidth}
            \includegraphics[width=\linewidth]{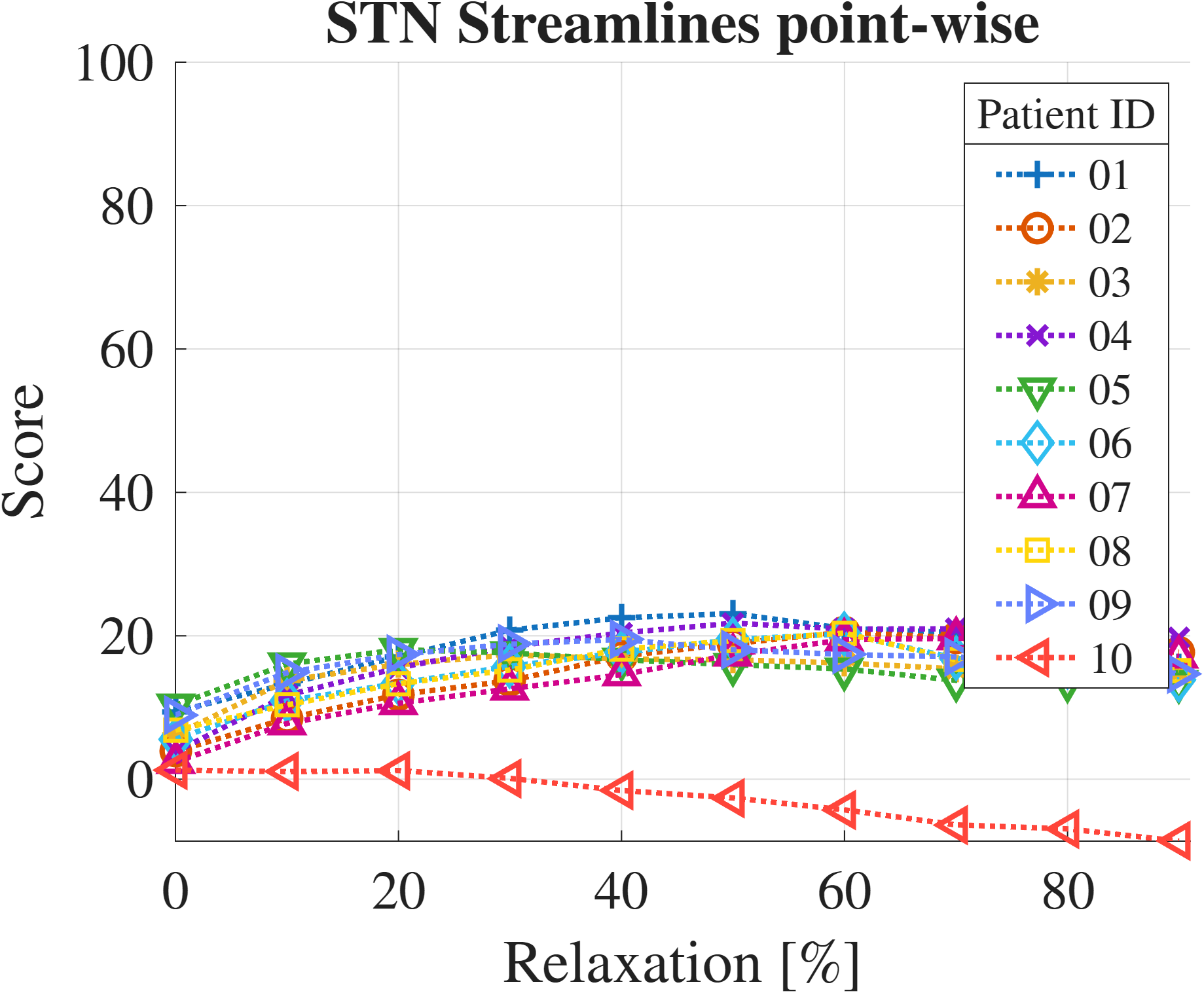}
        \end{subfigure}\vspace{0.5em}

        \begin{subfigure}[t]{\linewidth}
            \includegraphics[width=\linewidth]{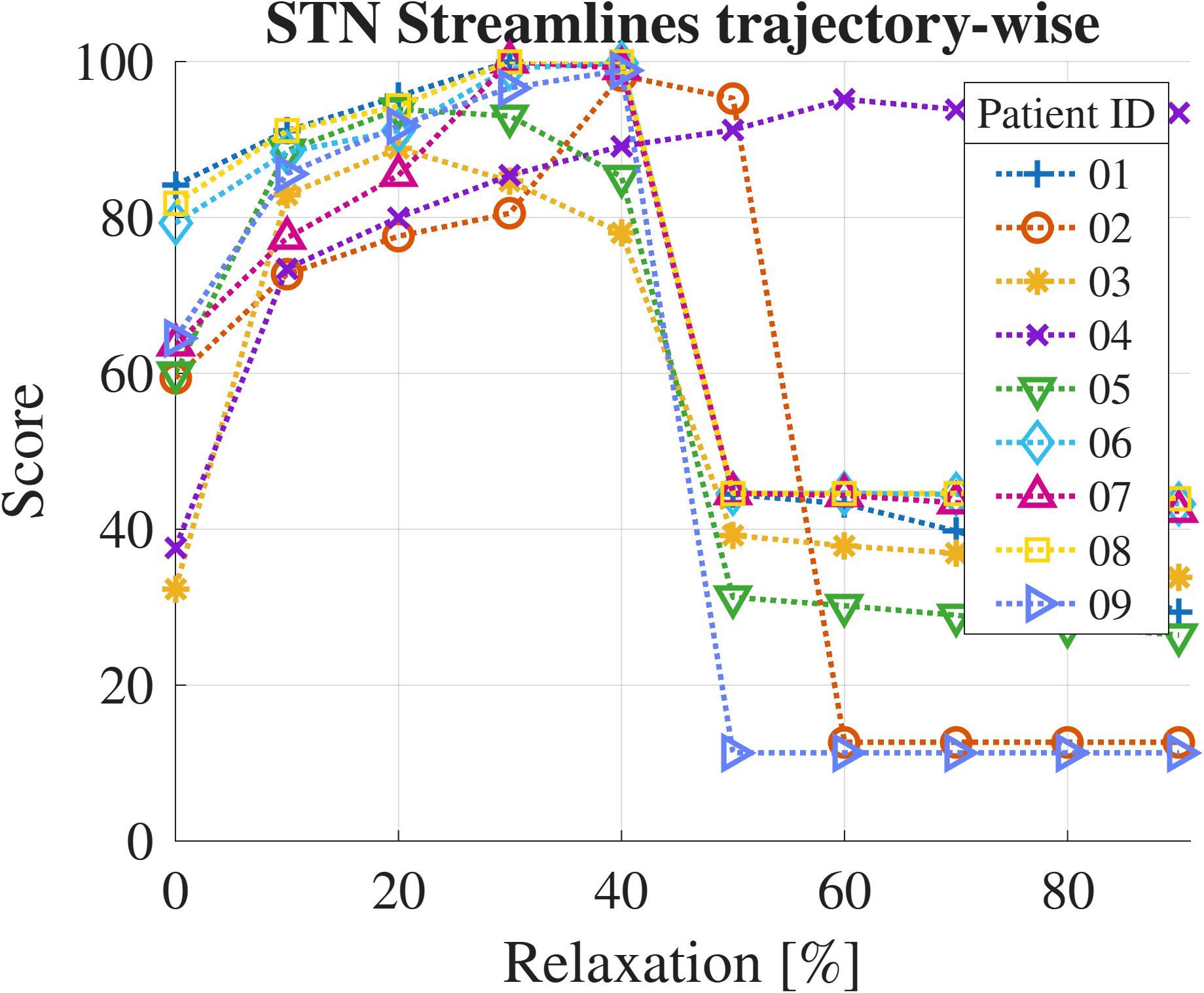}
        \end{subfigure}
    \end{minipage}
    \caption{Effect of constraint relaxation on the ranking score given by~\eqref{eq:score} for the linear optimization scheme in~\eqref{eq:opti1} and the nonlinear optimization scheme in~\eqref{eq:opti2}. Results shown for the right hemisphere only (left hemisphere in supplementary material). Peak scores are typically achieved at approximately 40\% relaxation across most patients. Lower scores are observed for the point-wise approach when targeting STN streamlines, as these contain more points distributed along the fiber length compared to the spatially compact STN subdivisions. }
    \label{fig:score} 

\end{figure}
\subsection{Optimal contact configurations}
\begin{figure*}[ht]
    \centering
    \begin{minipage}[c]{0.04\textwidth}
        \centering
        \rotatebox{90}{\textbf{Linear}}
    \end{minipage}%
    \begin{minipage}[c]{0.95\textwidth}
        \centering
        \begin{subfigure}[t]{0.32\linewidth}
            \includegraphics[width=\linewidth]{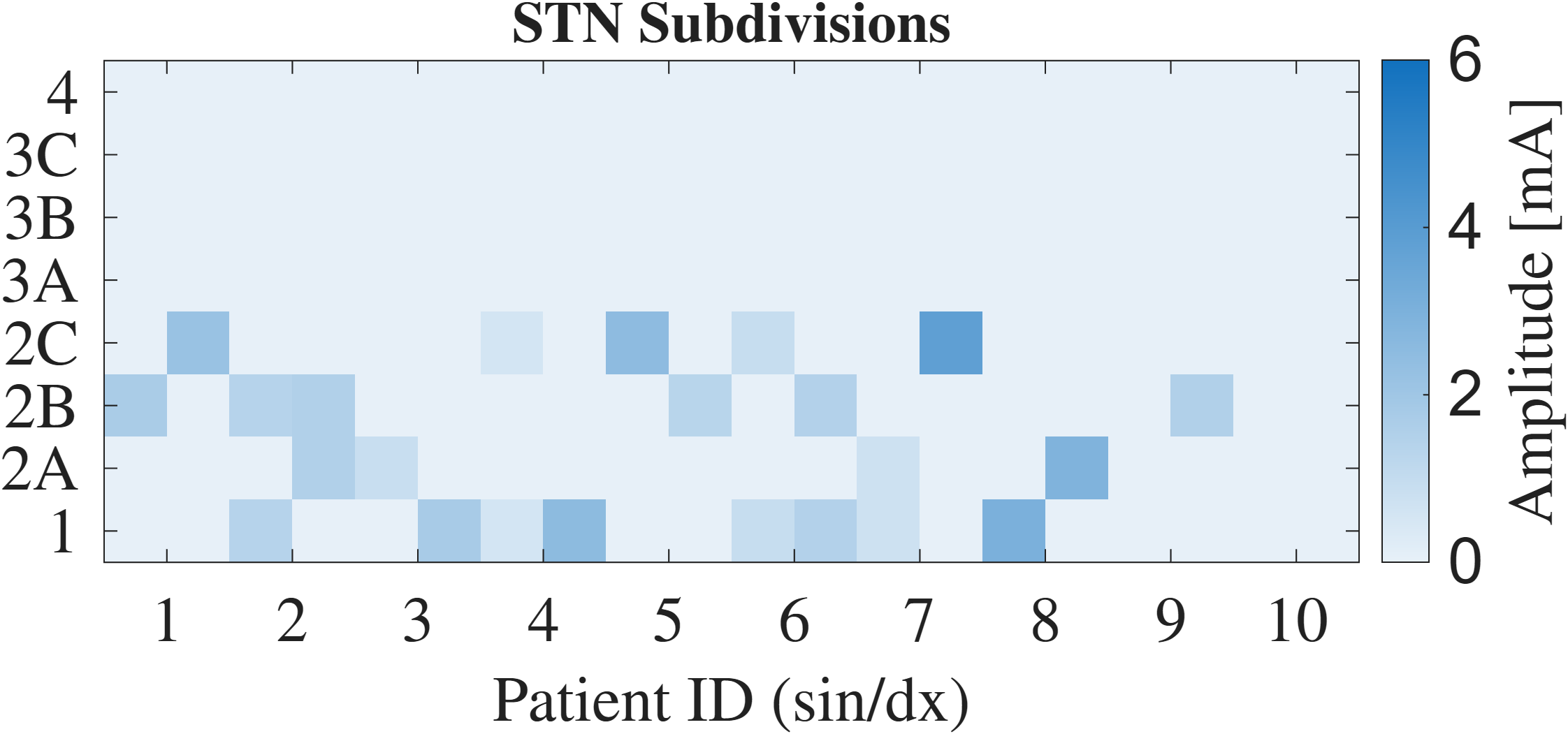}
        \end{subfigure}\hfill
        \begin{subfigure}[t]{0.32\linewidth}
            \includegraphics[width=\linewidth]{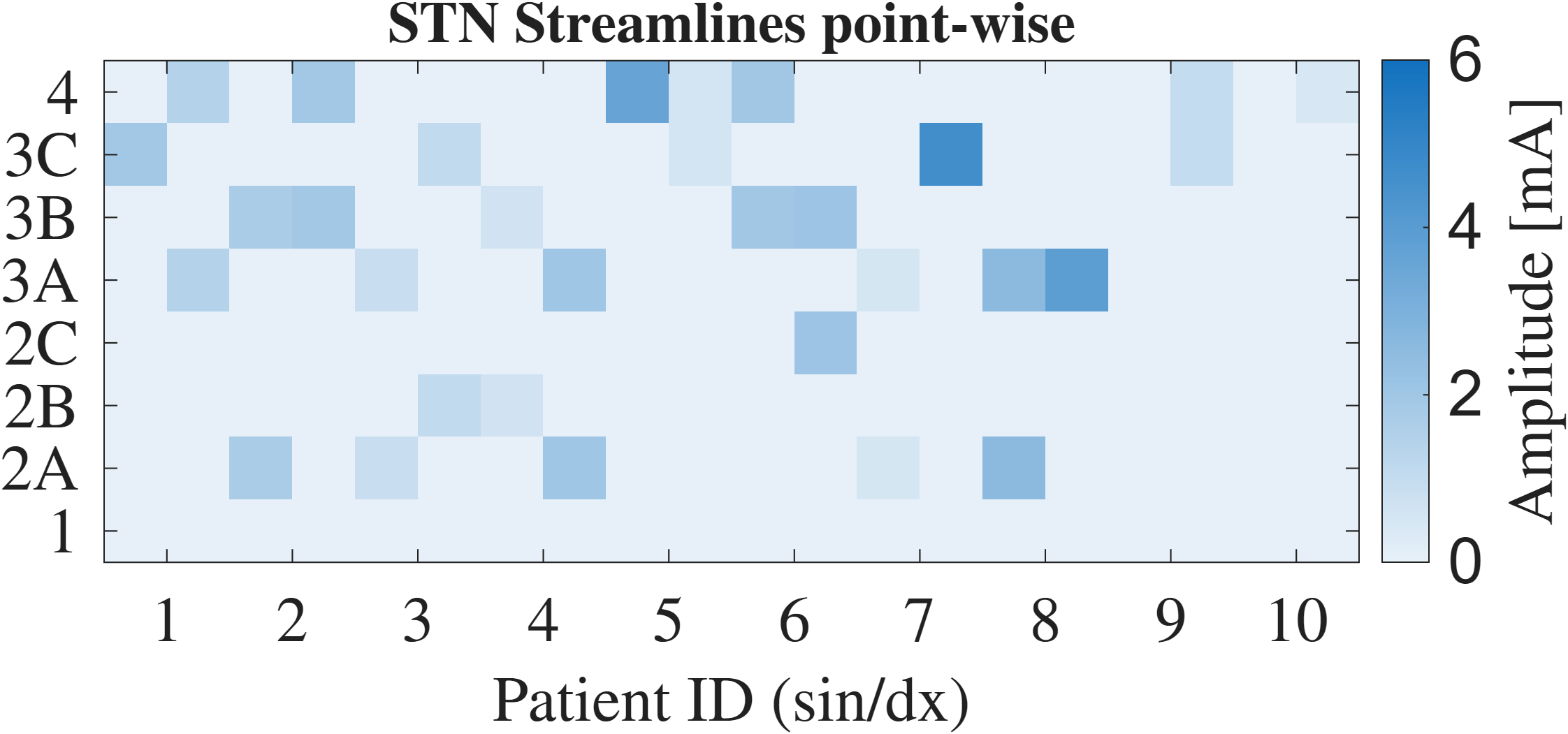}
        \end{subfigure}\hfill
        \begin{subfigure}[t]{0.32\linewidth}
            \includegraphics[width=\linewidth]{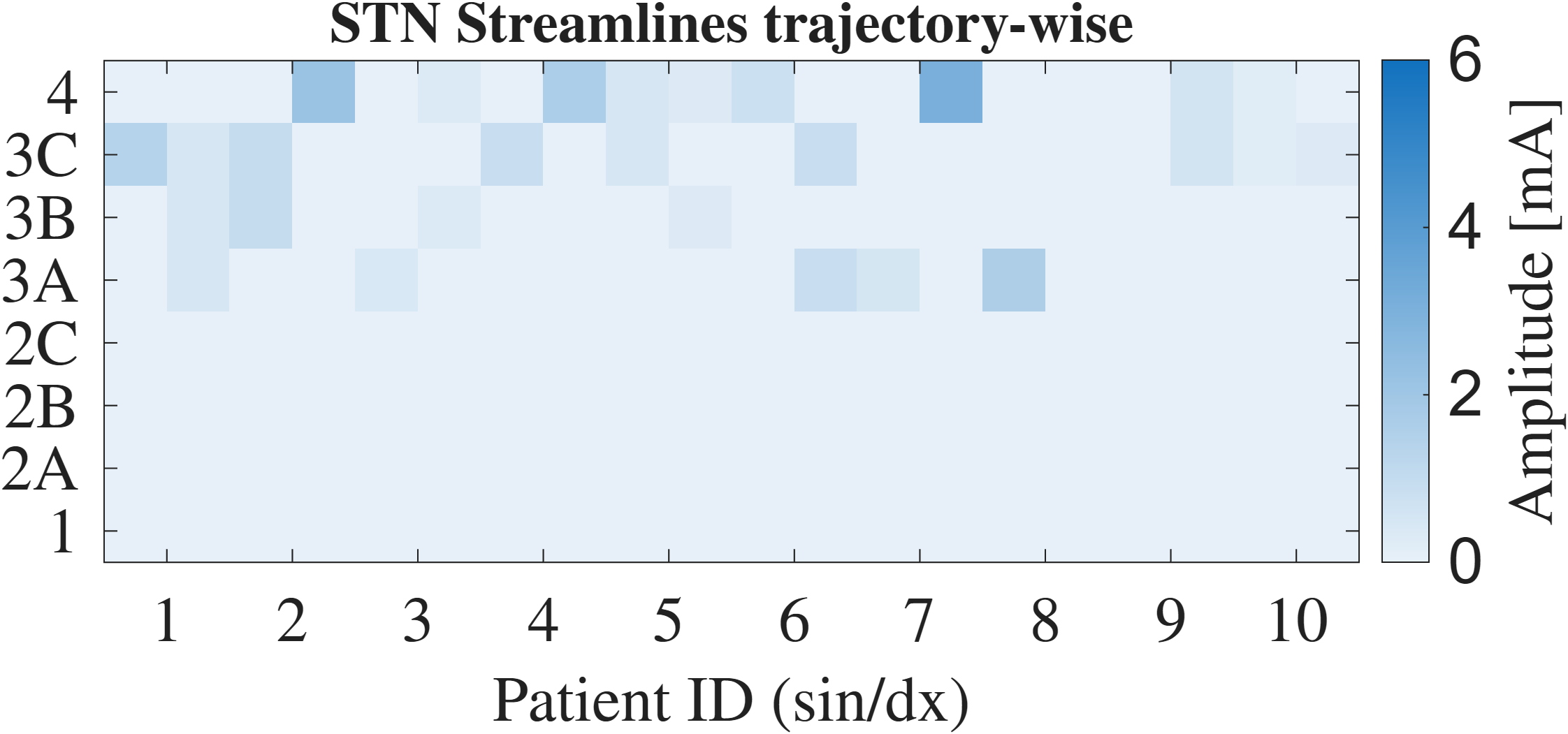}
        \end{subfigure}
    \end{minipage}
    
    \vspace{1em}
    
    \begin{minipage}[c]{0.04\textwidth}
        \centering
        \rotatebox{90}{\textbf{Nonlinear}}
    \end{minipage}%
    \begin{minipage}[c]{0.95\textwidth}
        \centering
        \begin{subfigure}[t]{0.32\linewidth}
            \includegraphics[width=\linewidth]{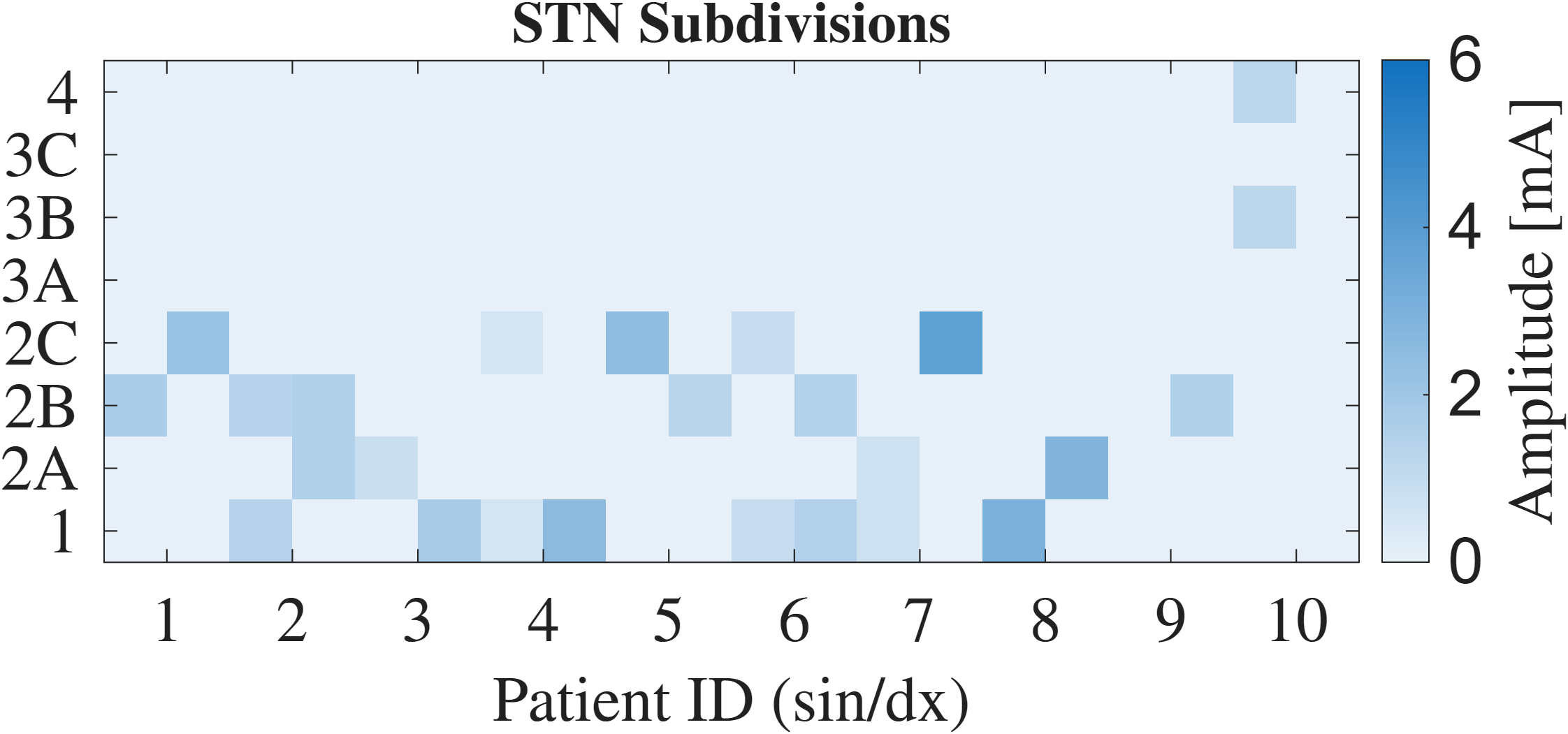}
        \end{subfigure}\hfill
        \begin{subfigure}[t]{0.32\linewidth}
            \includegraphics[width=\linewidth]{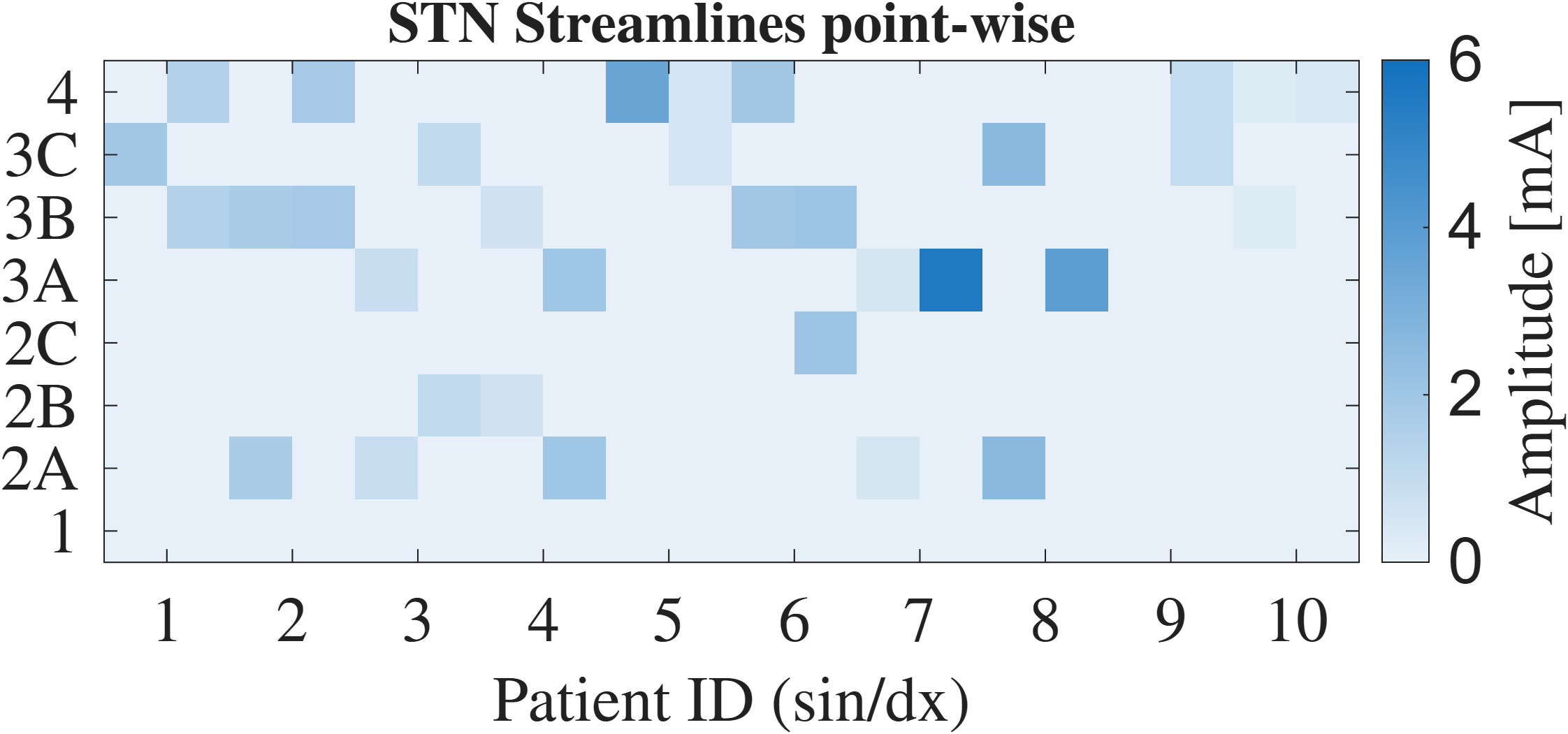}
        \end{subfigure}\hfill
        \begin{subfigure}[t]{0.32\linewidth}
            \includegraphics[width=\linewidth]{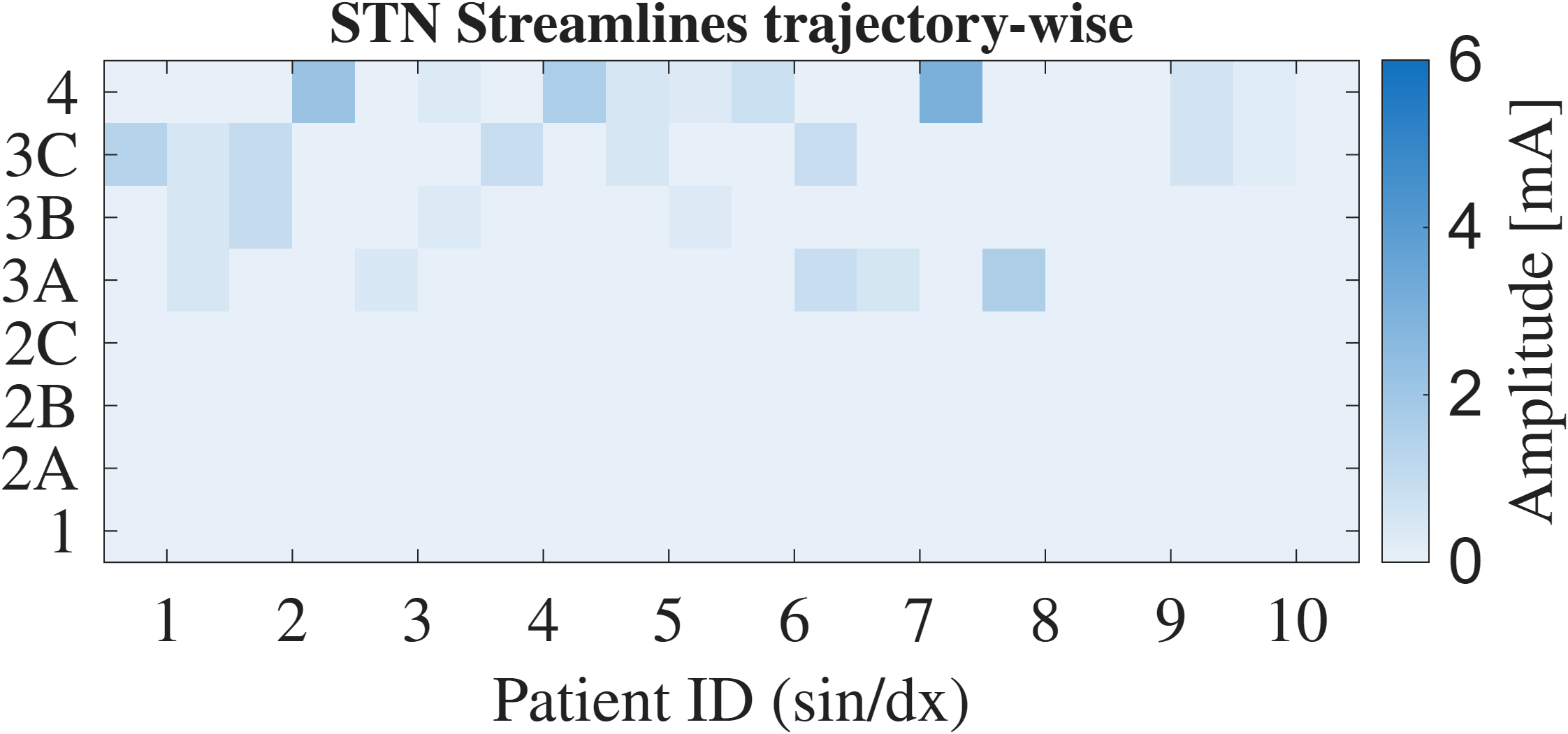}
        \end{subfigure}
    \end{minipage}
    
    \caption{Highest ranked stimulation settings suggested by the linear optimization scheme in \eqref{eq:opti1} and the nonlinear optimization scheme in \eqref{eq:opti2}, across all patients. The amplitude distribution across all contacts ranges from zero (white) to \SI{6}{mA} (dark blue). The optimized settings favor lower contacts (toward the tip of the DBS lead) when targeting the STN subdivisions, whereas contacts positioned higher are favored when targeting the STN streamlines.}
    \label{fig:optimal_suggestions}
\end{figure*}

The following section presents the contact configuration suggestions produced by solving optimization problems~\eqref{eq:opti1} and~\eqref{eq:opti2} for the individualized models in TuneS. 
The results presented in Fig.~\ref{fig:score} and Fig.~\ref{fig:optimal_suggestions} indicate that the two optimization schemes yield similar results, when applied to the same set of targets and constraints. Although the linear algorithm is quicker, both optimization schemes are efficiently solvable on a standard computer.

In the following, the sensitivity of the optimal solutions to variations in the constraint relaxation variable $\gamma$, as well as the impact of the chosen targets and constraints on the predicted configurations, are explored in more detail.

\paragraph{Robustness analysis} 
The effect of the constraint variable $\gamma$ on ranking  score \eqref{eq:score} is given in Fig.~\ref{fig:score}. 
Both the linear and nonlinear optimization schemes yield comparable results, achieving similar scores and showing consistent behavior for the individual DBS leads. A notable score reduction occurs at larger $\gamma$ values in the trajectory-wise approach, likely because portions of constraint streamlines are recruited together at higher relaxation levels and amplitudes.

\paragraph{STN subdivisions vs streamlines}
The top suggested settings that were ranked highest with respect to~\eqref{eq:score} for the linear and the nonlinear optimization scheme are illustrated in Fig.~\ref{fig:optimal_suggestions}.
The results highlight that targeting the STN streamlines, as opposed to its subdivisions, results in more recommendations for dorsally positioned contacts (higher along the $z$-axis) as well as more combinations involving vertically adjacent contacts. 
Further, it can be observed that much lower scores for targeting the streamlines compared to the STN subdivisions are reported in Fig.~\ref{fig:score}. This is a result of the point-wise computation of target coverage in the optimization scheme. Across most patients and optimization schemes, peak scores occurred at approximately 40\% constraint relaxation. However, both hemispheres in Patient 10 consistently yielded the lowest scores (often zero amplitude on all contacts), and the nonlinear trajectory-wise approach failed to converge for this patient. The point-wise approach for STN streamlines showed the most consistent performance across relaxation levels, indicating reduced sensitivity of the overall score to constraint relaxation compared to other target-scheme combinations.

\subsection{Comparison \textit{in silico}}
\begin{figure*}[t]
     \centering
    \begin{subfigure}[b]{0.32\linewidth}
        \centering
        \includegraphics[width=\textwidth]{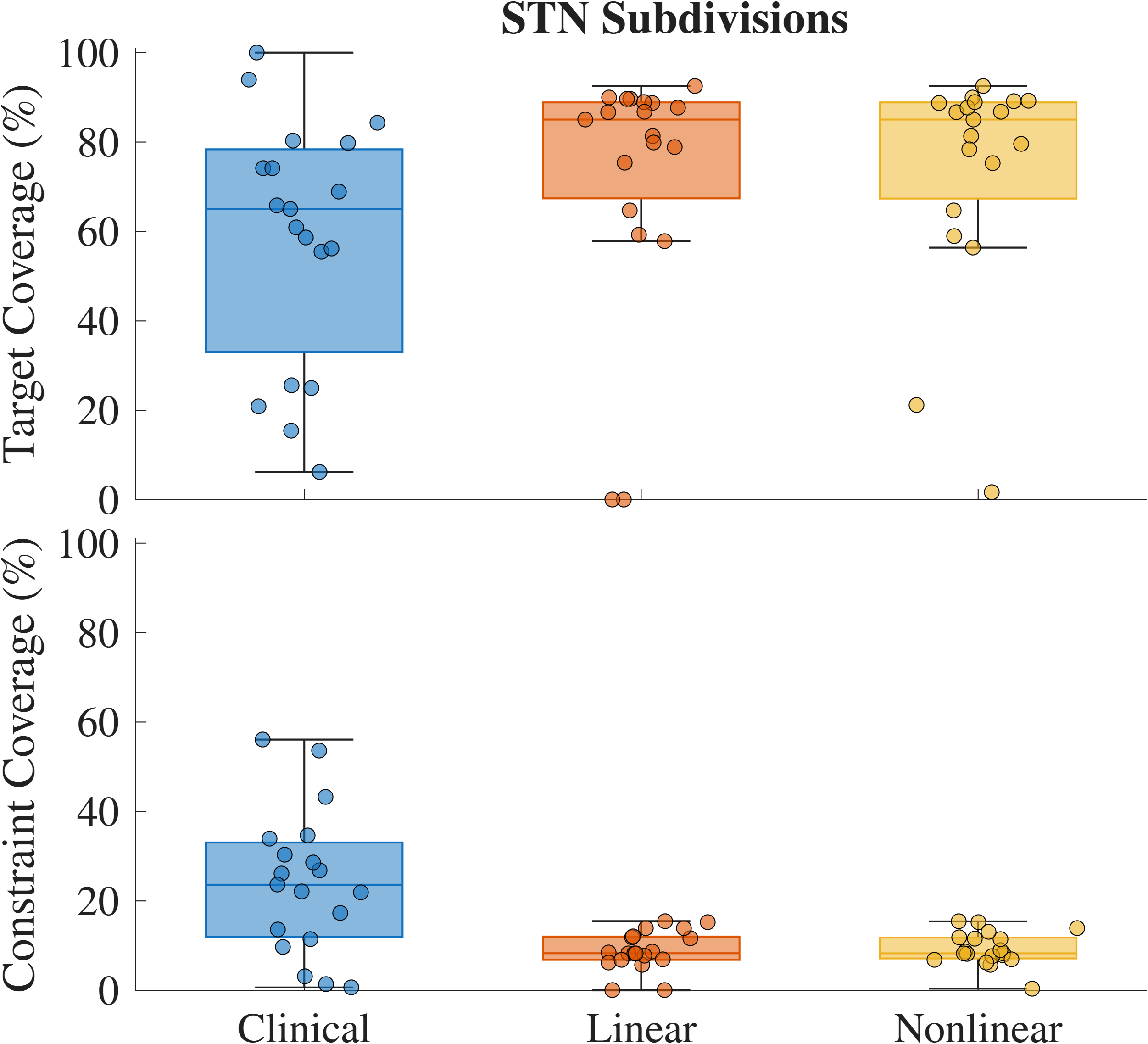}
    \end{subfigure}
    \begin{subfigure}[b]{0.32\linewidth}
        \centering
        \includegraphics[width=\textwidth]{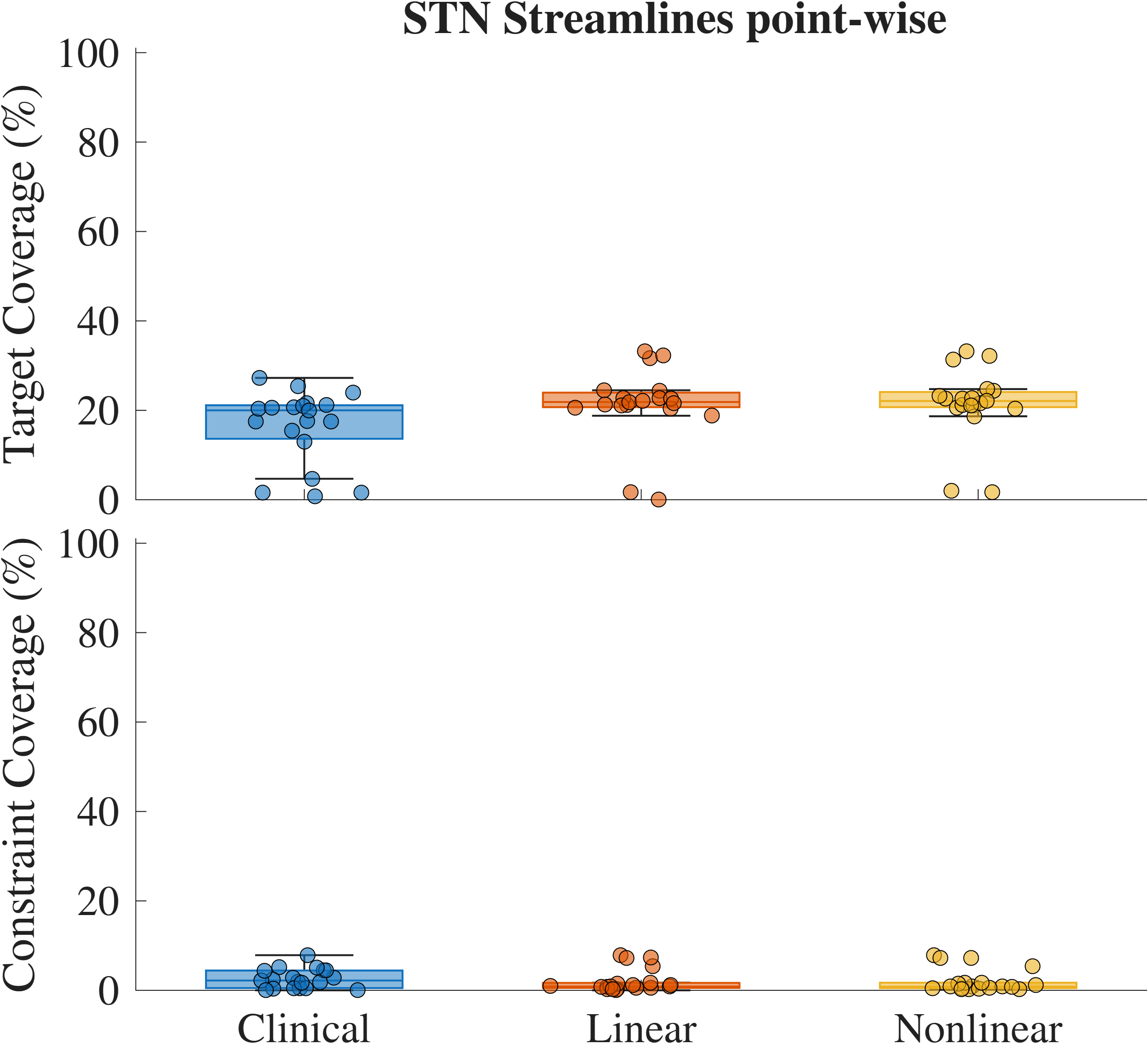}
    \end{subfigure}
    \begin{subfigure}[b]{0.32\linewidth}
        \centering
        \includegraphics[width=\textwidth]{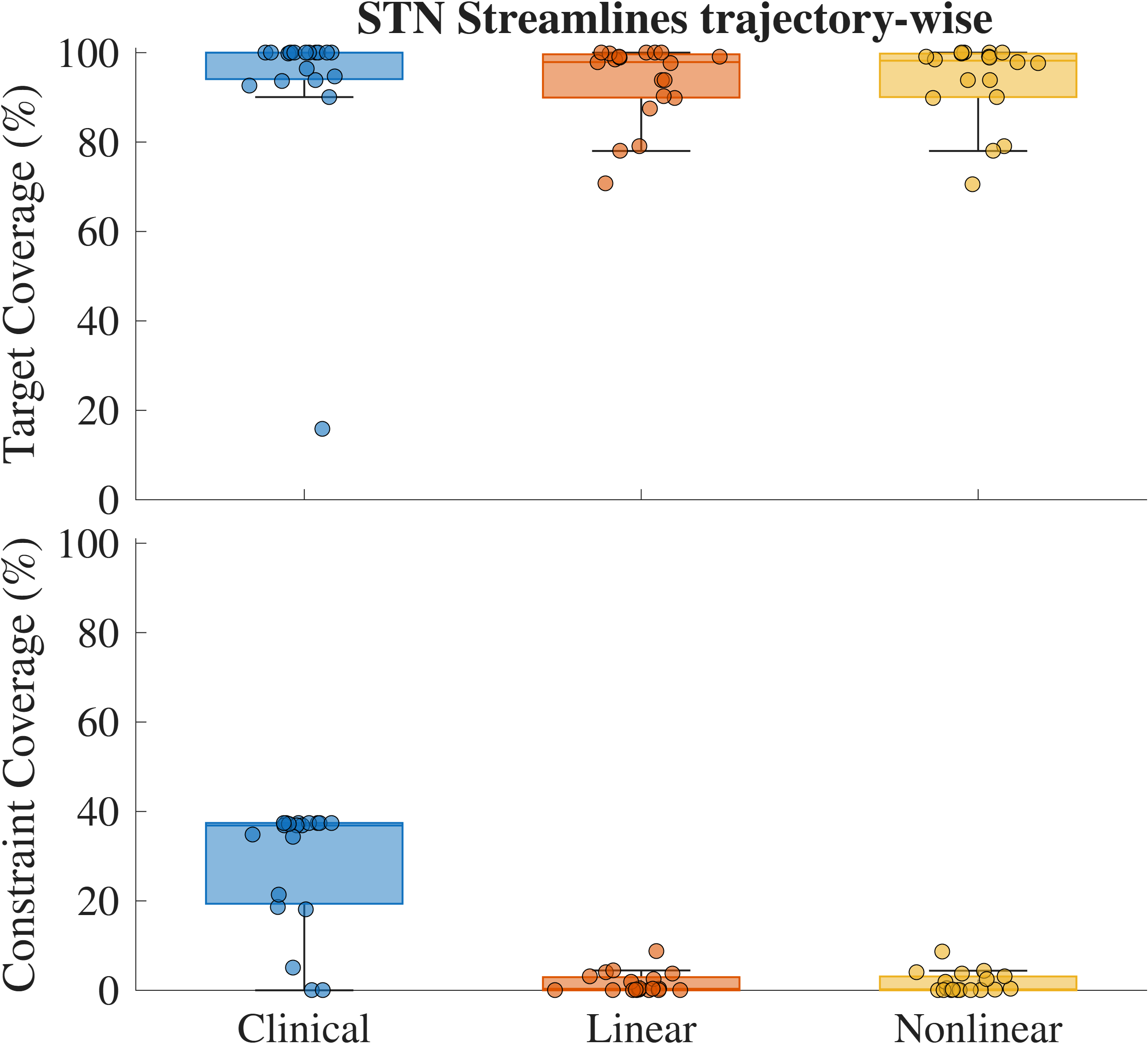}
    \end{subfigure}
    
    \caption{Cohort-level comparison of target and constraint coverages across different targets (STN subdivisions and STN streamlines) and activation approaches (point-wise and trajectory-wise). Optimized settings consistently achieve comparable or higher target coverage while reducing constraint activation compared to clinical settings.}
    \label{fig:Cohort_coverages}
\end{figure*}

The contact and amplitude predictions by TuneS are based on a fixed pulse width and frequency, corresponding to the selected VTA threshold. For a retroperspective comparison \textit{in silico}, the TuneS contact suggestions were compared to the clinically active settings using the clinically active amplitude and pulse width adjusted electric field norm thresholds~\cite{Astrom2015}. The cohort-level results for target coverages and constraints are illustrated in~Fig.~\ref{fig:Cohort_coverages}.

When targeting the STN subdivisions, the model-based predictions achieved higher target coverage while maintaining constraint coverage comparable to the clinical settings. A similar observation was made for point-wise targeting of the STN streamlines. However, in this case, the predicted settings required higher amplitudes than those clinically applied, as the constraint regions were not effective in limiting stimulation. Thus, the increased target coverage was primarily driven by higher amplitudes. For trajectory-wise targeting of the STN streamlines, the clinical settings achieved consistently high target coverage, whereas the optimized settings showed greater variability. Here, the automated algorithms aims to find a balance between maximizing target engagement and minimizing constraint activation, which resulted in much lower stimulation amplitudes as well as lower constraint coverages overall. If desired, the current trade-off could be replaced by an optimization algorithm that optimizes for full target coverage while minimizing the activation of constraints.

As illustrated in Fig.~\ref{fig:score}, Patient~10 consistently achieved the lowest scores compared to the rest of the cohort.  These findings align with the challenges faced during clinical programming sessions for this patient, which required frequent adjustments, and can be attributed to the lead's position, which is relatively far from the atlas-based target structure, as depicted in Fig.~\ref{fig:pat10}(b). Notably, an apparent shift of the lead position relative to the target can be observed between the CT scan acquired three days post-surgery and the one obtained several months later (Fig.\ref{fig:pat10}(b)). This effect is commonly attributed to brain shift, which may arise from factors such as air entry or loss of CSF during surgery. The observed difference in the reconstruction of the lead positions can reflect small lead movements as well as inaccuracies in the co-registration of pre- and post-operative images. The later CT scan was used to computed predicted settings and coverages.

\section{Discussion}

The results presented in this paper  are related to only one of the potential neuromodulation applications of TuneS. The pipeline is however not limited to a single disorder and can be employed to explore stimulation targets in a wide variety of those. 

\subsection{Clinical settings}
Clinically used settings typically provide satisfactory symptom relief, yet they do not necessarily constitute ground-truth of the optimal configuration. Moreover, clinical settings typically do not elicit any side effects and, therefore, are not expected to 
impose on constraint regions rather than achieve optimal target coverage. Thus, proving the therapeutical superiority of one contact configuration over another requires clinical validation supported by symptom quantification. In PD, objective quantification of symptoms and stimulation induced side-effects remains a challenge but is, for the time being, addressed by e.g. smartphone and smartwatch applications, eye tracking, and motion detectors~\cite{Olsson2020,Amprimo2024,Jansson2015}. 

\subsection{Optimal contact configurations}
\paragraph{Robusteness analysis}
 In clinical settings, all three segmented contacts A, B, C in one row (see Fig.~\ref{fig:leads}) are often used, whereas the use of individual segmented contacts is usually explored when the former settings do not yield a satisfying effect or produce side effects. In contrast, the automated algorithm typically favors a single contact or a combination of two segmented contacts to avoid spill from contacts that are distant from the target.
It should be noted, that although just two optimization problems are investigated in this paper, additional cost functions for different targets and constraints are readily available or can be implemented in TuneS by the user.

Furthermore, different contact configurations often perform similarly in simulation and are therefore ranked closely. Therefore, as mentioned above, proving the efficiency of one combination over another requires clinical testing. 

\paragraph{STN subdivisions vs streamlines}
The choice of targets and constraints strongly influences the prediction of the optimal contact configuration. In particular, targeting the STN streamlines typically resulted in more dorsal contact predictions compared to the STN subdivisions, see Fig.~\ref{fig:optimal_suggestions}. The dorsal region of the STN and adjacent white matter have repeatedly been reported as a ``sweet spot" in STN DBS~\cite{Dembek2019, Roquemaurel2021}, leading to motor improvements in larger PD patient cohorts. Given the findings in the present paper, this suggests that STN streamlines may equally well be used in image-guided programming algorithms. This is in line with recent discoveries of disease- and symptom-specific ``sweet streamlines"~\cite{Hollunder2024, Rajamani2024},  which could be integrated in TuneS as targets in future modifications. In particular, symptom-specific streamlines offer potential to tailor the stimulation based on the patient's specific symptoms. 

Although the recommendations by TuneS for optimal contact configurations are generally consistent — often suggesting the same contacts or their immediate neighbors for given targets and constraints — the suggested amplitudes can vary depending on user-defined degrees of freedom in the optimization schemes. These degrees of freedom include the relaxation of the chosen optimization scheme, the VTA threshold, and the selection of the ranking coefficients, cf. \eqref{eq:score}. As previously reported by~\cite{Cubo2019}, safety constraints that produce amplitudes comparable to those used clinically are highly patient-specific and may be influenced by individual factors like tissue conductivity, stimulation sensitivity, and symptom severity. The dynamical stimulation parameters, i.e. pulse length and frequency, also could play a role in the variability.

\subsection{Limitations}
The results obtained with TuneS are subject to several limitations, which the user should be well aware of. 

First, a simple static model is used for modeling the VTA produced by the DBS lead. The static model does not take into account dynamic parameters such as frequency and pulse width. Further, it was previously shown that VTAs for bipolar configurations are inadequately represented by VTAs computed from electric field norm thresholds~\cite{Duffley2019}. Consequently, bipolar settings were not treated in this paper. 
Further, potential differences in tissue conductivity between patients were neglected by assuming the same conductivity values for GM, WM, and CSF in all patients. 

Second, TuneS incorporates the lead orientation obtained from post-operative CT scans, yet the CT scans were taken for clinical purposes and may not always provide the optimal features for the orientation detection algorithm. Depending on the slice thickness and the angle between the slice and the lead axis, reconstructing the lead orientation based on the DBS marker can be difficult. Determining whether the orientation is pointing in one direction or the opposite direction (180 degrees away) is particularly challenging due to the symmetry of the marker artifact. This uncertainty can significantly affect the suggestion of a specific contact segment. However, new photon-counting CT scanners may offer the possibility to very clearly and unambiguously determine lead orientation~\cite{Manfield2024}.

Third, the brute-force approach of testing a given number of contact combinations limits computational efficiency, particularly as the number of individual contacts increases with more advanced lead designs. Moreover, this study only considers evenly distributed current across all active contacts, whereas modern pulse generators allow for uneven current distributions, offering greater flexibility. 

Finally, this study explored the inclusion of streamlines as both targets and constraints within automated optimization schemes. We compared point-wise and trajectory-wise activation approaches, showing that the point-wise method often led to unrealistically high amplitudes due to insufficient constraints. In contrast, the trajectory-wise approach provided a more intuitive representation of white matter tract engagement and facilitated higher target coverage. Future research may benefit from incorporating disease- or symptom-specific ``sweet streamlines''~\cite{Hollunder2024,Rajamani2024} as targets, together with corresponding ``sour streamlines'' as constraints, to achieve more conclusive results.
It is important to note that streamline atlases typically use probabilistic mappings derived from large patient cohorts. As a result, while these streamlines may exhibit general similarities, they do not necessarily match the patient-specific anatomical fibers, and individual responses to stimulation are likely to vary.

A version of TuneS with reduced dependencies, particularly COMSOL, is a work in progress to enable more users to incorporate TuneS into their workflow.
Currently, only atlas-based targets are supported out of the box, but manually defined targets and constraints can be incorporated by the user.

\section{Conclusion}
This paper introduces TuneS, an automated optimization pipeline for DBS settings, demonstrating its value as a research tool for exploring various target and constraint regions. Although the findings presented here are based on a small cohort of Parkinson's disease patients, TuneS holds promise for application across a range of neurological and mental disorders. Notably, it extends automated optimization algorithms to the targeting of streamlines, which have been at a matter of increasing research interest. Nevertheless, several open questions remain, particularly regarding the quantification of streamline activation in a static model and whether the current model complexity is sufficient to explain patient outcomes. To address these challenges, future efforts should focus on integrating simulation results with objective symptom quantification measurements.

\section*{Acknowledgment}
The computations were enabled by resources provided by the National Academic Infrastructure for Supercomputing in Sweden (NAISS) and the Swedish National Infrastructure for Computing (SNIC) at UPPMAX partially funded by the Swedish Research Council through grant agreements no. 2022-06725 and no. 2018-05973.

\bibliographystyle{IEEEtran}
\bibliography{IEEEabrv,references}

\end{document}